\documentclass[pre,twocolumn,showpacs,floatfix,nofootinbib,groupaddress,showkeys,superscriptaddress]{revtex4}
\usepackage{graphicx}
\usepackage[dvips]{color}
\usepackage{epsfig}
\usepackage{amssymb}
\usepackage{amsmath}
\usepackage{bm}

\begin{document}

\bibliographystyle{apsrev}
\title{Radial Domany-Kinzel Models with Mutation and Selection}
\author{Maxim O. Lavrentovich}
\email{mlavrent@physics.harvard.edu}
\affiliation{Department of Physics, Harvard University, Cambridge, Massachusetts 02138, USA}
\author{Kirill S. Korolev}
\affiliation{Department of Physics, Massachusetts Institute of Technology, Cambridge, Massachusetts 02139, USA}
\author{David R. Nelson}
\affiliation{Department of Physics, Harvard University, Cambridge, Massachusetts 02138, USA}
\begin{abstract}

We study the effect of spatial structure, genetic drift, mutation, and selective pressure on the  evolutionary dynamics in a simplified model of asexual organisms colonizing a new territory.  Under an appropriate coarse-graining, the evolutionary dynamics is related to the directed percolation processes that arise in voter models, the Domany-Kinzel (DK) model, contact process, etc.  We explore the differences between linear (flat front) expansions and the much less familiar {\it radial} (curved front) range expansions. For the radial expansion, we develop a  generalized, off-lattice DK model that minimizes otherwise persistent lattice artifacts.  With both simulations and analytical techniques, we study the survival probability of advantageous mutants, the spatial correlations between domains of neutral strains, and the dynamics of populations with deleterious mutations.  ``Inflation'' at the frontier leads to striking differences between radial and linear expansions.    For a colony with initial radius $R_0$ expanding at velocity $v$, significant genetic demixing, caused by local genetic drift, only occurs up to a finite time $t^* = R_0/v$, after which  portions of the colony become causally disconnected due to the inflating perimeter of the expanding front.  As a result, the effect of a selective advantage is amplified relative to  genetic drift, increasing the survival probability of advantageous mutants.  Inflation also  modifies the underlying directed percolation transition, introducing novel scaling functions and modifications similar to a finite size effect. Finally, we consider radial range expansions with \textit{deflating} perimeters, as might arise from colonization initiated along the shores of an island.    

\end{abstract}

\date{\today}

\maketitle

\section{\label{SIntro}Introduction}

To grow and prosper, populations must often migrate into new territory. These ubiquitous range expansions occur  in bacterial growth on a Petri dish, tumor growth, viral infections, human migration, species movement induced by climate change, and in many other biological systems \cite{KorolevBac, cancerEvo, virus1, virus2, iceageEvo, humanmigration, climate1, climate2, frogs, worms}.   Such expansions influence the evolutionary dynamics of the population with, for example, enhanced genetic drift due to low population densities at the frontier.   To understand the universal features of these diverse range expansions,  it is of interest to construct simple models with the essential elements of the population's evolutionary and spatial dynamics.

Many features of a range expansion can be captured by adapting a simple ``stepping stone'' model \cite{kimurapaper, KorolevRMP}.
In this model, individuals in a collection of island subpopulations, or demes,  reproduce, die, and mutate stochastically.  Two commonly used stochastic processes are the Moran and Wright-Fisher models (see \cite{KorolevRMP, Moran, Ewens} and references therein).     The demes track spatial configurations, which simulate the  spatial distribution of an evolving population.   By allowing individuals to migrate between adjacent demes, one can simulate the effect of  dispersal on the evolutionary dynamics in, say, one or two dimensions.   Analysis of these models highlighted many important features distinguishing the evolutionary dynamics of a well mixed  versus  spatially extended  populations  (see \cite{KorolevRMP} for a review). 

As discussed in Ref.~\cite{KorolevRMP}, stepping stone models also provide a useful approximation to population genetics at the {\it frontier} of an expanding population.  First, consider an asexual population expanding across a flat surface with an approximately linear front.  Under a wide variety of conditions,   the front will move with a constant speed $v$, set by the competition between growth and dispersal (see e.g., Ref.~\cite{murray} and references therein).  This growth is called a Fisher wave \cite{fisher}.    Second, we assume that only individuals living near the population front are able to divide and settle into the new territory.  In the approximation of a {\it flat} population front, and in a reference frame moving with the frontier, we can simulate the front using a one-dimensional stepping stone model.    This model predicts that for small deme size, number fluctuations (genetic drift) create a strong spatial genetic demixing of the population into domains of closely related individuals.  Initially, demixing occurs primarily through local fixation within a deme. At long times, however, coarsening is dominated by the diffusive motion of the domain boundaries, which annihilate or coalesce on contact, creating larger domains.  Selective advantages will bias the motion of the boundaries, and mutations will introduce new domains into the population \cite{KorolevRMP}. 
 \begin{figure}[!ht]
\includegraphics[height=2.1in]{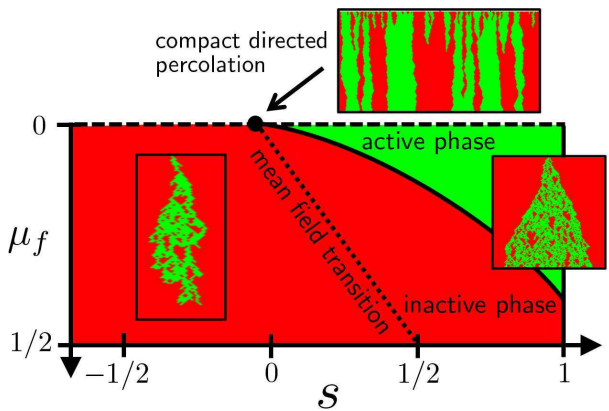}
\caption{\label{fig:PhaseDiagram}  (Color online) A sketch of the phase diagram for mutation-selection balance in linear expansions with irreversible, deleterious mutations from green (light gray) to red (dark gray) cells occurring at rate $\mu_f$, at a fitness cost $s$ (see Appendix~\ref{AMutSelBal} for a biological motivation).  The  solid line between red (dark gray) and green (light gray) regions represents the directed percolation phase transition.  We show as insets  typical clusters generated from single green (light gray) cells in the active and inactive phases. The dashed line at the top represents the case of no mutations where the system falls into the compact directed percolation (CDP) universality class.  The dynamics at criticality ($s = \mu_f = 0$ at the black dot) in this CDP regime is  shown.  The dotted line shows the position of the mean field transition (i.e., for a well mixed population).  An analogous plot for inflating radial range expansions is shown in Fig.~\ref{fig:DKPhase}.}
\end{figure}

Stepping stone models (and the simplified voter-type models discussed below) of population genetics at frontiers make use of the notion of ``dimensional reduction.''  A range expansion of organisms across a surface takes place in two spatial and one temporal dimensions.   At the cost of considerable computational complexity, a full $2+1$-dimensional model of the population genetics could be constructed, as was done recently within mean field theory to model range expansions of baker's yeast, {\it Saccharomyces cerevisiae} \cite{KorolevMueller}. However, the population density and relative proportions of many organisms stop changing (and some even stop migrating) in the wake of a front colonizing virgin territory.  The crucial dynamics of genetic drift, competition, and spatial diffusion then takes place at a linear frontier, where one can approximate the $2+1$-dimensional dynamics by a $1+1$-dimensional stepping stone in a reference frame which moves with the population front.

This mapping is only exact for perfectly {\it flat} population fronts. For neutral mutations, the flat front approximation leads to boundaries of genetic domains behaving as simple random walks which wander from their initial positions with a characteristic displacement $\Delta x \propto \sqrt{ t}$ that increases with the square root of time.  Interesting complications arise when the front undulates -- for neutral genetic variants, the boundaries between genetic domains couple to the undulations and wander more vigorously, which changes some of the detailed predictions of the theory \cite{Saito, DRNPNAS, KorolevRMP, nelsonhallatschek}.  We neglect such complications here, but note that Kuhr et al. \cite{frey} have recently studied dynamical phase transitions in a two-species Eden model with an approximately flat but undulating front, and found {\it different} critical exponents than those of the directed percolation universality class \cite{Hinrichsen} considered here.  It's worth noting that not only are flat one-dimensional habitats easier to analyze, a quasi-one-dimensional environment could describe the evolutionary dynamics along the bank of a river, along a seacoast, or along a constant altitude slope of a linear mountain range.  To obtain inflation (or deflation) in this context, one could invoke, for example, a receding or advancing waterline around an island.  Flat front models might also describe the gradual, climate-driven shift of some populations toward the Earth's poles    \cite{climate1, climate2}.

An interesting special case of the stepping stone model is one in which a fit wild-type strain undergoes deleterious mutations that confer a fixed selective disadvantage.  If the deleterious mutations are irreversible, there will be a resultant selection-mutation balance, as shown in Fig.~\ref{fig:PhaseDiagram}.  The irreversible, deleterious mutations describe the genetic dynamics of a well adapted population evolving in a fitness landscape with a single sharp peak (see Appendix~\ref{AMutSelBal} for more details).   The population sector dynamics exhibits a phase transition that falls into the directed percolation universality class \cite{Hinrichsen}.  The order parameter is the fixation probability of the advantageous strain.  In the ``active'' phase, the advantageous strain has a finite probability of surviving after an infinite amount of time (in a spatially infinite population).  In the ``inactive'' phase, the strain always dies out eventually.  The transition into the inactive phase is sometimes called ``mutational meltdown'' in population genetics literature \cite{nelsonhallatschek,mutmeltdown}.  We consider the simplest possible model leading to directed percolation: a single active strain with deleterious mutations.  However, directed percolation behavior is observed in a wide variety of biological processes  \cite{virus1,virus2,firedfire}.  

In this paper we study \textit{curved} population fronts in two dimensions that expand with some constant velocity $v$, motivated by populations that settle new territory from some  localized, approximately circular, initial homeland with radius $R_0$. (A semicircular variant of this theory would describe organisms spreading into the interior after a coastal colonization event.)  A particularly simple example is the growth of a circular  bacterial inoculation on a Petri dish \cite{hallatschekPNAS}.  We will  again assume that the  dynamics are confined to the population front.  A dimensional reduction to a $1+1$-dimensional model will also be possible in this  case, but the spatial dimension must now \textit{inflate} to account for the growth of the curved front. To simplify the analysis, we assume the extreme limit of a stepping stone model, with a single cell per deme.  As a result, the genetic drift, or number fluctuations, at the frontier will be very strong. We will argue  that the long-time, large-size dynamics of  this extreme limit of the stepping stone model is equivalent to the dynamics of a generalized voter model under an appropriate coarse-graining.

In general, the curved front suppresses the effects of genetic drift by systematically increasing the perimeter of the frontier and inflating sector boundaries away from each other.  The suppression becomes significant after a crossover time $t^* = R_0/v$.  A scaling argument for this crossover time follows from a comparison of the diffusive and inflationary length scales in the problem.  At short times, sector boundaries move diffusively and sectors grow to a characteristic  diffusive  size $\ell_d \sim R_0 \phi \sim  \sqrt{D t}$, where $\phi$ is the angular sector size, $R_0$ is the homeland radius, and $D$ is a diffusion constant.  The diffusion constant scales as $D \sim a^2/\tau_g$  where $a$ is the cell diameter, and $\tau_g$ is the time between generations (see Sec.~\ref{SRadial} for more details).   The inflationary growth due to the increasing population radius $R(t) = R_0 +vt$ has a characteristic size $\ell_i \sim \phi vt$.  So, there will be a crossover time $t^*$ at which the two length scales $\ell_i$ and $\ell_d$ are comparable.        This sets $\ell_d\sim R_0 \phi \sim \sqrt{D t^*} \sim \phi v t^* \sim \ell_i$ and we find the desired result  $t^* \sim R_0/v$. After time $t^*$, sector boundaries with a significant angular separation no longer diffusively interact, and the population separates into independent segments.  Consequently, the survival probability of an advantageous mutant become enhanced relative to linear expansions, because the advantageous mutant just has to survive until time $t^*$ after which inflation prevents its extinction via genetic drift.

The inflationary dynamics also strongly influences the mutation-selection balance (directed percolation phase transition) by introducing a cutoff to the critical population dynamics  after time $t^*$.   Just as in a finite size effect at a conventional phase transition, this cutoff is characterized by  scaling functions.  We find that the dynamics scales with $t^*$ in a similar way as for a finite population size $L$ in a linear range expansion at criticality, but with a completely different scaling function.  Moreover, inflation is qualitatively different from typical finite size effects due to transient dynamics or small system sizes \cite{finitesize,Hinrichsen}:  Inflation occurs at long times and is due to a gradually \textit{increasing} system size. 

We will also consider the effects of inflation on the dynamics of single sectors of cells with a selective advantage.  A key dimensionless parameter that characterizes the strength of the selective advantage for voter-type models (in the presence of inflation) is $\kappa \approx s \sqrt{R_0/2v \tau_g} $, where $s \leq 1$ is a selection coefficient, $\tau_g$ is the time between generations, and $v$ is the circular front propagation speed.  For $\kappa \ll 1$, selection is weak, genetic drift dominates the evolution, and the sector boundaries exhibit strong fluctuations.  Conversely, when $\kappa \gg 1$, the selective advantage is strong, the sector boundaries move out deterministically at long times, and the fate of the sector is decided very early in the evolution.  Examples of both cases are shown in Fig.~\ref{fig:SingleSeed}.   

In general, if a subpopulation has a strong selective advantage, it will create ``bulges'' in an advancing population front.  The population front will also  roughen due to the stochastic nature of the growth at the interface \cite{KorolevRMP, KorolevMueller}.  These bulges and  undulations will couple the population dynamics to the front geometry, changing the nature of the dynamics.     As discussed above, these effects are not included in our study.  However, we note that a smooth front approximation has been successfully employed in   studies of  expansions of bacteria, such as {\it Pseudomonas aeruginosa} \cite{KorolevBac}. In addition, this simplification will allow us to make connections between the radial expansions studied in this paper and the large body of work on stepping stone models for linear expansions with uniform fronts \cite{KorolevRMP}.

The organization of the paper is as follows: In Sec.~\ref{SSimulations} we develop a generalized Domany-Kinzel model  \cite{domany} to simulate both linear and radial range expansions. An unfortunate artifact in lattice simulations of circular range expansions is genetic boundaries and colony shapes that lock into the discrete four- or six-fold symmetry of the underlying lattice.  We describe novel simulation methods (which create an amorphous packing of organisms) that resolve this problem.  In Sec.~\ref{SRadial} we interpret the simulation results by appealing to a generalized voter model \cite{voter1,  redner} description that includes both mutations and natural selection. We also discuss how our generalized voter model is connected to the stepping stone model and the directed percolation universality class.
The effect of inflation  on the directed percolation phase transition is also described in Sec.~\ref{SRadial}. In addition, we compare linear and radial expansion results for quantities such as fixation probabilities and the heterozygosity correlation function.   Sec.~\ref{SSReverseInflation} presents a brief discussion of an alternative ``deflating''  inward expansion in which the population front gets smaller, as might happen in colonization from the perimeter of an island.  We conclude  with final observations and a summary in Sec.~\ref{SConclusions}.

\section{\label{SSimulations}Simulations}

Many properties of the directed percolation phase transition line shown in Fig.~\ref{fig:PhaseDiagram} have been studied using the Domany-Kinzel (DK) lattice model \cite{domany, Hinrichsen, lubeck}.   The DK model is implemented on a hexagonal (i.e. equilateral triangle)  lattice  in which  horizontal lines of sites  are updated one at a time, starting from the top.  The sites can be in one of two states (e.g., red or green) and the states of newly evolved sites are determined by the states of neighboring sites in the horizontal line above, as shown in  Fig.~\ref{fig:rules}$(a)$.  Since all of the updating occurs at the bottommost actively evolving horizontal line, we have the dimensional reduction discussed in the introduction.  The downward direction on the lattice can be treated as a time axis.  

The DK model has a natural biological interpretation.  The sites represent cells, and the evolving lines of sites represent cell generations evolving as flat population fronts. Note that the DK model updates the entire front or generation of cells before moving the front.  This scheme is called parallel updating \cite{NEQPTBook, Hinrichsen} and has non-overlapping generations.  The two states of the cells in this model correspond to two genetic variants or ``alleles.''   During each update step, new ``daughter'' cells are evolved according to a probabilistic rule that  depends on the states of  neighboring ``parent'' cells in the previous generation.  Although these  parents do not sexually reproduce, two neighboring asexual cells must compete for space for daughter cells during reproduction.   Thus,  not all of the parent cells will be able to propagate an associated daughter cell because of limited space at the frontier.  As part of this reproductive process, successful daughters are allowed a small chance of switching colors via mutation, as discussed below.  After a daughter cell is pushed forward into a frontier populated by a new generation, cell division behind the advancing frontier ceases.

\subsection{\label{SSDK} A Generalized Domany-Kinzel Model}

\begin{figure}
\includegraphics[height=1.6in]{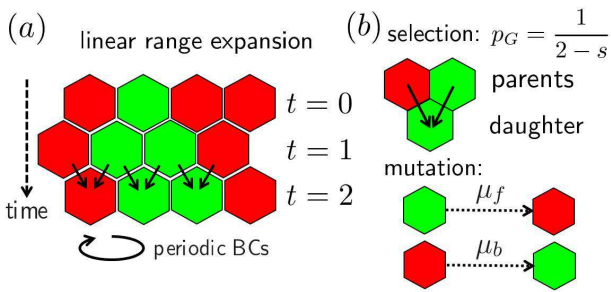}
\caption{\label{fig:rules} (Color
  online) $(a)$ A sketch of the  Domany-Kinzel (DK) model implemented on a hexagonal lattice so that each daughter cell has two potential parents.  Periodic boundary conditions are employed, and each cell is evolved using input from the states of the two cells above it in the previous generation. $(b)$ Illustration of the two update steps in our generalized DK model.     The daughter shares the color of two parents of the same color. Otherwise, the daughter is  green (light gray) with probability $p_G=(2-s)^{-1}\approx 1/2+s/4$ if its parents are different colors, where $s \leq 1$ is the selective advantage of the green strain.    We set $p_R = 1-p_G= \frac{1-s}{2-s} \approx 1/2-s/4$.  A newly evolved cell is then allowed to mutate from green (light gray) to red (dark gray) with a forward mutation probability $\mu_f \ll 1$ or vice-versa with a backward mutation probability $\mu_b \ll 1$.                            }
\end{figure}

In the original version of the DK model, there are two parameters $p_1$ and $p_2$ which set the update rules \cite{domany}.  If a daughter cell has a  green and red parent in the previous generation, the daughter is green with probability $p_1$.  If both parents are green, the daughter is green with probability $p_2$.   When the parents are both red, the daughter cell is always red so that no mutations from red to green cells are allowed.  This 
 parameterization is not well suited for population genetics, so a biologically motivated modification that includes two-way mutations is necessary.
 To adapt and extend the DK model, daughter cells are now created in two steps shown in Fig~\ref{fig:rules}$(b)$.  In the first step, a green or red daughter cell is created with a probability $p_G$  or $1-p_G$, respectively.  We construct $p_G$ by drawing inspiration from the Moran model, a well studied update scheme for well mixed populations \cite{Moran,murray}.  The probability $p_G$ will be proportional to both the number of potential green parents $n_G$, and the green cell growth rate, which is normalized to unity.  The red cells  suffer a selective disadvantage and grow at the smaller rate  $1-s$.  Then, to ensure that  $p_G$ is properly normalized, we have
\begin{equation}
p_G = \frac{n_G}{n_G+n_R(1-s)}=\frac{n_G}{ sn_G +n_T(1-s)}, \label{eq:pG}
\end{equation}  
where $n_R$ is the number of red parents and $n_T \equiv n_R+n_G$.   After this first selection step, the surviving daughter cell mutates forward with probability $\mu_f\ll 1$ if it is green and  backward with probability $\mu_b \ll 1$ if it is red.  The update steps are shown for a simple linear expansion  in Fig.~\ref{fig:rules}$(b)$, for which $n_T=2$.     

\begin{figure}
\includegraphics[height=2in]{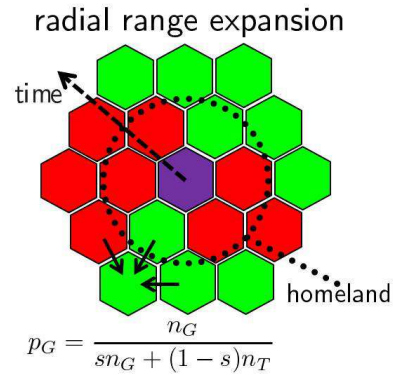}
\caption{\label{fig:Updates}  (Color
  online) An illustration of a radial expansion simulation on a hexagonal lattice with an initial state given by a ``homeland'' of  7 cells.  We add daughter cells one at a time, each time picking an unoccupied lattice site that is closest to the center of the homeland (the central purple cell).  When two or more cells are at the same distance, we chose one of them at random.  We implement the generalized Domany-Kinzel update rule that allows for more than two potential parent cells.  Each daughter cell will be green (light gray) with a probability $p_G$ that will depend on the number of  green (light gray) parents $n_G$, given by Eq.~(\ref{eq:pG}).  Mutations can be implemented in a second update step as described in the text (also see Fig.~\ref{fig:rules}).                               }
\end{figure}

  A  lattice model for radial range expansions can now be constructed by taking advantage of the variability of $n_T$ in the update rule in Eq.~(\ref{eq:pG}).  For example, as illustrated in Fig.~\ref{fig:Updates}, we can use a hexagonal lattice and evolve the cells starting from an initial homeland centered around a reference cell.  We always choose to evolve the daughter cell that is closest to the central reference cell in the initial seed.  This choice ensures a uniform, circular population front.  Certain daughters will now have three $(n_T=3)$ instead of two $(n_T=2)$ parents.  Our revised DK update scheme easily adapts to this variation.  As discussed below, this algorithm leads to circular colonies, and might arise biologically if previously deposited cells generate a chemical that stimulates cell division at the frontier.

\subsection{\label{SSBennett} The Bennett Model}

Unfortunately, the \textit{radial} DK model implemented on lattices with discrete rotational symmetries (e.g. hexagonal, square, or triangular lattices)  exhibits strong lattice artifacts.  For example,  domains of cells prefer to grow along the crystallographic directions, which can introduce artificial oscillations in angular correlation functions, as discussed in Sec.~\ref{SSMutations}.      To eliminate this problem, we have implemented the DK model on an amorphous lattice that does not have any special directions.  An efficient way to generate such a lattice is provided by an extension of the Bennet algorithm \cite{Bennett}.

 In the original Bennet model  \cite{Bennett}, identical hard sphere particles (cells, in our case) were deposited sequentially in locations closest to a chosen reference point that becomes the center of the population.  The possible locations for daughter cells are generated by enumerating all the possible ways of placing a new cell so that it touches at least two  cells without overlapping any other  cell.   The first few close-packed cells (a triangle of three, in the planar implementation discussed below) are placed manually in an initial ``seed''.  It turns out that, in two dimensions, the generated cluster of identical cells has a very strong hexagonal ordering, regardless of the shape of the initial seed \cite{rubinstein}.   Following Ref.~\cite{rubinstein}, we create an amorphous configuration by depositing cells with two different sizes with different probabilities.  For a carefully chosen range of ratios $\rho$ between the smaller and larger cell radius and   the probability $Q$ of depositing the smaller cell rather than the larger one, one finds packings with no preferred directions.

\begin{figure}
\includegraphics[height=3.1in]{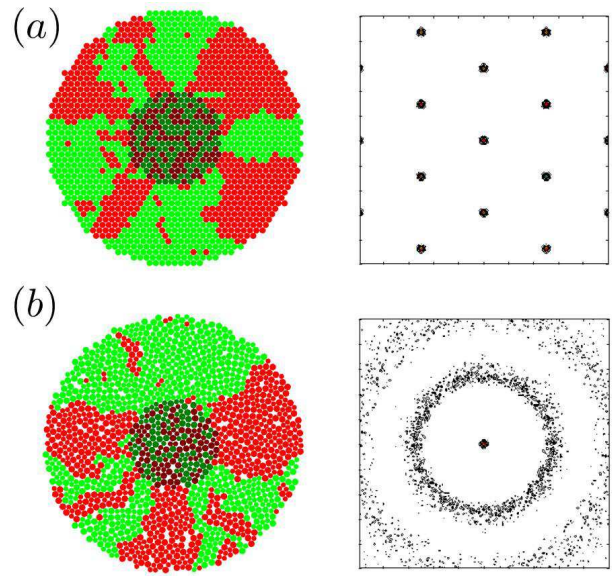}
\caption{\label{fig:structure} (Color
  online) $(a)$ A radial DK model simulation on a hexagonal lattice.  The structure factor of the lattice is shown in the right panel.  The cell colors were assigned with neutral selective advantage $(s=0)$ and no mutations $(\mu_f=0)$.  The first 1500 cells generated by the algorithm are shown.  The central shaded region is the homeland of radius $R_0 \approx 10$ cell diameters, where  green (light gray) and  red (dark gray) cells are placed randomly with equal probability.  $(b)$ The same simulation performed on a Bennett model lattice with  approximately equal numbers of small and large cells $(Q~=~0.5)$ and a radius ratio $\rho = 0.769$. The structure factor in the right panel shows that the Bennett cell packing is isotropic.}
\end{figure}

  One test of rotational symmetry for a generated cluster is to compute the structure factor
\begin{align}
S(\mathbf{k}) \equiv \frac{1}{N} \left| \sum_{i=1}^N e^{i \mathbf{k} \cdot \mathbf{r}_i} \right|^2,
\end{align} 
where $N$ is the number of cells in the cluster and $\mathbf{r}_i$ is the center of the $i$-th cell.
A hexagonal cluster of cells with identical radii will have a striking structure factor with a clear six-fold symmetry, as shown in Fig.~\ref{fig:structure}$(a)$.  By contrast, the amorphous cluster in Fig.~\ref{fig:structure}$(b)$ has a structure factor with no discrete symmetries.    The latter cluster was generated using equal numbers of large and small cells $(Q~=~0.5)$ and size ratio $\rho = 0.769$.    We use this choice of parameters in the radial range expansion simulations in the next two sections (Sec.~\ref{SSNeutralEvo} and Sec.~\ref{SSMutations}).  In the later sections, we use slightly different parameters ($Q = 0.6$ and $\rho = 8/11 \approx 0.73$) which yield similar cell configurations.

Once an amorphous configuration of cells is generated, a homeland of cells is established by assigning genotypes of red or green to all cells within a radius $R_0$ of the central reference point.  Fig.~\ref{fig:structure}$(b)$ illustrates a well mixed initial homeland with $R_0$ equal to about ten cell diameters.  This initial homeland is then evolved using the rules discussed in Sec.~\ref{SSDK}.  Similar to the evolution on the hexagonal lattice in Fig.~\ref{fig:Updates}, a new daughter is always the one that is closest to the central reference point used to generate the lattice.  This can be regarded as an approximation for cell colonies where previous generations continuously secrete a chemical promoting cell division into the open space at the frontier.  (We assume steric constraints prevent cell division in the interior of the colony.)  The resulting concentration of stimulant will  monotonically decrease from the center, leading to circular colonies.  Alternatively, a circular population front can be imposed externally by, for example, a receding waterline around a circular island.  The time $t$ is measured in generations.  Each generation represents the time needed to evolve a new rim of cells at the circular population front.   We assume that this rim has a finite thickness corresponding to the average cell width, so that the population radius is given by $R(t) = R_0 + vt$.  The speed $v$ is  the rim thickness divided by the generation time.

To facilitate the measurement of time in units of the radial colony size, we rescale all of our amorphous lattice coordinates such that we get the same cell number density as in a hexagonal lattice of identical cells with a diameter equal to one.  We know that for the hexagonal lattice, a radial expansion will grow out by about a cell diameter per generation.  Thus, with this rescaling, we can treat each generation in our amorphous lattice as all the cells in a rim with a thickness equal to one.  This means that our population front speed is  set to $v = 1$ and the radius of the radial range expansion at time $t$ is now  $R(t) = R_0 + t$.   

In the Bennett model lattice, there are a variety of ways to choose the competing parents in the local neighborhood of each daughter.  In our simulations, we choose the parents to be the previously evolved cells that touch the daughter cell.  Since we update the cells one at a time, starting from the closest cell to the reference point, the adjacent parent cells must all be closer to the reference point than the daughter.  This usually gives us two or three potential parents.  However, sometimes the daughter cell is adjacent to cells that are all further from the reference point and have not yet been updated.  In  these exceptional cases, we  choose the competing parents to be the two closest nonadjacent cells that are closer to the reference point.

Our amorphous cell packings will have slight inhomogeneities.  For example, there will be a variation in the coordination number $z$ at each lattice site.   The lattice spacing $a$ will also vary across the lattice.  Hence, we define an effective coordination number $z_r$ for our radial expansions as well as an effective lattice constant $a_r$.  Since the  amorphous lattice  is not the densest possible packing of the cells, there will be gaps between the cells, and $a_r$ is expected to be slightly larger than the average cell diameter.   In particular, we expect the amorphous lattice to be more loosely packed than a hexagonal lattice of identical cells with diameters equal to the average cell diameter in the amorphous lattice.

The inhomogeneities, or ``quenched disorder,'' in our amorphous cell packings require an average over spatial variations to get the effective parameters $a_r$ and $z_r$.  We have found it most straightforward to estimate $a_r$ and $z_r$ by fitting simulation results to known analytic solutions. For example, the heterozygosity function derived in Sec.~\ref{SSNeutralEvo} is used to fit the combination $a_r^2/z_r$ (see Fig.~\ref{fig:hetcollapse}).   An alternative approach might be to use the coarse-graining procedure used to produce measured correlation functions from actual cell colonies in Ref.~\cite{KorolevBac}.  Some quantities, such as the fraction of green cells $f(t)$ at time $t$,  can be measured without knowing either $a_r$ or $z_r$.   Note that the disorder is present in both the time-like (radial) and the space-like (circumferential) direction on the lattice.  Fortunately, it can be shown \cite{Hinrichsen} that spatio-temporally quenched disorder is an irrelevant perturbation in systems in the directed percolation universality class and it will not change the important features of the dynamics.

\section{\label{SRadial} Voter Models with Inflation}

To understand the radial simulation results analytically, we shift to a generalized voter model description in which cells in the population are randomly updated one at a time, instead of generation by generation.  The former and latter update schemes are often referred to as random-sequential and parallel updates, respectively \cite{NEQPTBook}.   The original Domany-Kinzel model uses parallel updates as discussed previously.  Our simulations (using the DK model generalization  discussed in Sec.~\ref{SSimulations}) represent an intermediate case since we evolve the system one cell at a time, but in a deterministic way to ensure that the generations are updated one after another. The random-sequential scheme is used in canonical models such as the voter model or more general contact processes, which have  features conducive to an analytical analysis  \cite{NEQPTBook}.  Parallel updates are easier to implement in simulations.  Despite the differences between these two update schemes, we expect that they share universal features when we look at their coarse-grained dynamics.  There are exceptional cases where this is not so, such as in the roughening transition in models of polynuclear growth \cite{Hinrichsen}.   However, we do not expect our simulations to be exceptional, because the directed percolation transition can be modelled with either update scheme \cite{Hinrichsen}.  We start be considering flat frontier range expansions, and then introduce the inflation due to radial range expansions in the coarse-grained description at the end.

To start the analysis, we exploit dimensional reduction and consider the cells at the frontier of a $d+1$-dimensional population as ``voters'' in a $d$-dimensional space.  It is convenient to describe the states of the cells with a set of Ising spin variables $\{ \sigma_i \} _{i \in \mathbb{Z}^d}$ where $\sigma_i = \pm 1$  and $i$ is a $d$-dimensional vector of integers describing the location of the $i^{\mathrm{th}}$ cell on a hypercubic lattice that approximates cell positions in our amorphous simulations.  The spin variable encodes the cell state: $\sigma_i = 1$ for a green cell and  $\sigma_i = -1$ for a red one.  No empty sites are allowed, which is equivalent to assuming a  uniform front that advances in lock-step in $(d+1)$-dimensions with each generation.  We now will use the notation $\{ \sigma \} \equiv \{ \sigma_i \}_{i \in \mathbb{Z}^d}$ for convenience.  The probability distribution of a particular spin configuration $\{ \sigma \}$, $ P \left(\{ \sigma \},t \right)$, obeys the master equation
\cite{redner}
\begin{align}
\partial_t P(\{ \sigma \},t)  & =  \sum_{j \in \mathbb{Z}^d}  \left[ \omega_{  \{  \sigma\}_j \rightarrow \{ \sigma \} } P(\{ \sigma\}_j,t) \right. \nonumber \\[5pt]
&  \qquad \qquad \qquad\left. - \omega_{\{ \sigma \} \rightarrow \{  \sigma\}_j} P(\{\sigma\},t) \right],
\label{eq:masterequation}
\end{align}
where  $\{  \sigma\}_j$ is the same configuration of spins as $\{ \sigma \}$, \textit{except} that the $j^{\mathrm{th}}$ spin is flipped (has an opposite state).  Spin flipping in the biological context represents the replacement of a cell at site $j$ by a cell of the opposite type after cell division.  The variable $j$ in Eq.~(\ref{eq:masterequation}) runs over all  spin locations and $\omega_{\{ \sigma \} \rightarrow \{  \sigma\}_j}$ are the probability rates of transitions between the configurations $\{ \sigma \}$ and $ \{ \sigma\}_j$.    The flipping rates $\omega_{\{ \sigma \} \rightarrow \{ \sigma\}_j}$ are given by the sum of three terms,
\begin{equation}
\omega_{\{ \sigma \} \rightarrow \{  \sigma\}_j}=\omega_{\{ \sigma \} \rightarrow \{  \sigma\}_j}^{\mathrm{drift}}+\omega_{\{ \sigma \} \rightarrow \{  \sigma\}_j}^{\mathrm{mut}}+\omega_{\{ \sigma \} \rightarrow \{  \sigma\}_j}^{\mathrm{sel}}, \label{eq:merates}
\end{equation}
where $\omega_{\{ \sigma \} \rightarrow \{  \sigma\}_j}^{\mathrm{drift}}$ is the rate due to random genetic drift, $\omega_{\{ \sigma \} \rightarrow \{  \sigma\}_j}^{\mathrm{mut}}$ is the contribution from mutations, and $\omega_{\{ \sigma \} \rightarrow \{  \sigma\}_j}^{\mathrm{sel}}$ is the contribution from a selection bias.

For neutral strains without mutations, only genetic drift contributes to the rates $\omega_{\{ \sigma \} \rightarrow \{ \sigma\}_j}$.   This genetic drift arises because a cell is more likely to change its state if it is surrounded by neighbors of the opposite type \cite{gardiner}, just as in the original voter model \cite{redner}.  Thus, $\omega_{\{ \sigma \} \rightarrow \{  \sigma\}_j}^{\mathrm{drift}}$ is the standard  rate  used in the voter model: 
\begin{equation}
\label{eq:fliprates}
\omega_{\{ \sigma \} \rightarrow \{ \sigma \}_j}^{\mathrm{drift}}= \frac{1}{2\tau_g} \left[1- \frac{\sigma_j}{z} \sum_{i \, \, \mathrm{n. \, n.}\, j } \sigma_i \right],
\end{equation}
where the summation is over the $z=2d$ nearest neighbors of voter $\sigma_j$. The generation time $\tau_g$ corresponds to updating every cell once per generation.
In one dimension $(d=1)$, Eq.~\ref{eq:fliprates} is equivalent to the zero temperature Glauber rate for the Ising model \cite{glauber}.   Mutations allow some cells to stochastically change their states independently of the states of their nearest neighbors.  We model these processes by introducing the rates
\begin{equation}
 \omega^{\mathrm{mut}}_{\{ \sigma \} \rightarrow \{ \sigma\}_j} =\frac{\mu_f(1+\sigma_j)}{2\tau_g} +  \frac{\mu_b(1- \sigma_j)}{2\tau_g},  \label{eq:mutfliprates}
\end{equation}
where $\mu_f$ and $\mu_b$ are the mutation probabilities for the cell during a generation time $\tau_g$.  These probabilities correspond to the mutation probabilities  in the DK model simulations described in Sec.~\ref{SSimulations}.   To make further contact with the DK model,  green cells are replaced by red cells with a rate that is smaller than the reverse rate by a factor of $1-s$, where $s \leq 1$ represents a selective advantage.  The resultant contribution to the flipping rate is
\begin{equation}
\omega^{\mathrm{sel}}_{\{\sigma \} \rightarrow \{ \sigma \}_j}= \frac{s}{4z\tau_g}\sum_{i \, \, \mathrm{n. \, n.}\, j }  (\sigma_i-1)\left(1+\sigma_j\right), \label{eq:selfliprates}
\end{equation}
where we sum over the nearest neighbors of spin $\sigma_j$.

We may now apply a coarse-graining procedure on the master equation (Eq.~(\ref{eq:masterequation})) with the rates in Eq.~(\ref{eq:merates}) to find a stochastic differential equation for the local, coarse-grained fraction $f(\mathbf{x},t)$ of green cells:
\begin{equation}
f(\mathbf{x},t) \equiv \frac{1}{\Omega} \sum_{i \in \mathcal{B}_{\Omega}(\mathbf{x})}  \frac{1+\sigma_i(t)}{2},
\end{equation}
where $\mathcal{B}_{\Omega}$  is a set of spins in a neighborhood  around frontier point $\mathbf{x}$ with volume $\Omega$.  In Appendix~\ref{ACoarseGrain} we derive that for $\Omega = 1$,  $f(\mathbf{x},t)$ obeys the equation 
\begin{align}
 \partial_t f  (\mathbf{x},t)&= D\, \nabla^2 f(\mathbf{x},t)+ \bar{s} \,f(\mathbf{x},t)[1- f(\mathbf{x},t)] \nonumber \\
& \quad  - \bar{\mu}_f f(\mathbf{x},t)+ \bar{\mu}_b \,[1-f(\mathbf{x},t)]+  \eta(\mathbf{x},t),  \label{eq:LangevinVMfull}
\end{align}
where $D = a^2/(z \tau_g)$ is the diffusion coefficient, $\bar{s} = s/\tau_g$, $\bar{\mu}_f=\mu_f/\tau_g$, $\bar{\mu}_b=\mu_b/\tau_g$, and  $\eta $ is a noise with correlations given by
\begin{align}
\left\langle\eta(\mathbf{x},t)\eta(\mathbf{x}',t) \right\rangle & \approx   \frac{2f(\mathbf{x},t)[1-f(\mathbf{x},t)]}{\tau_g} \nonumber \\
& \qquad \qquad \times  a^d \delta(t-t') \delta(\mathbf{x}-\mathbf{x}'). \label{eq:LangevinCorrVMfull}
\end{align}
This Langevin equation can also be derived starting from a stepping stone model where the lattice sites are  subpopulations undergoing a Moran process (see Appendix~\ref{AMutSelBal}) and which exchange individuals between neighboring sites.  By setting the number of individuals in each subpopulation to one, we find the same Langevin equation (Eq.~(\ref{eq:LangevinVMfull})), as shown in Ref. \cite{KorolevRMP}.   Because of the nonlinearity in the noise correlations Eq.~(\ref{eq:LangevinCorrVMfull}), the stochastic differential equation must be interpreted according to the \^Ito calculus \cite{gardiner,KorolevRMP}.  The time and space scaling used in the simulations sets $\tau_g = 1$.

The analysis so far is valid for both radial and linear range expansions as no reference has been made to the shape of the population front.  For a linear front of length $L$ in a two-dimensional population, the coordinate $\mathbf{x}$ is replaced by a single coordinate $x \in[-L/2,L/2]$.  For a circular front, we have to be more careful.  In an ``inflating'' front, the number $N$ of active ``voters''   or cells around the perimeter increases in time. To cover a variety of interesting cases, we let the radius $R(t)$ grow in time according to a general power law 
\begin{equation}
R(t) = R_0 \left[1 + \left( \frac{t}{t^*} \right)^{\Theta} \right], \label{eq:growingradius}
\end{equation}
where $\Theta$ is the growth exponent $0 < \Theta < \infty$, $t^*$ is a characteristic time, and $R_0$ is the initial homeland radius.  In our simulations, we will be restricted to $\Theta = 1$ where there is a constant front propagation velocity $v$ such that $t^* = R_0/v$.  The choice of $\Theta = 1$ is also appropriate for biological systems \cite{murray,palumbo,gray,hallatschekPNAS} expanding on a flat substrate.   

\begin{figure}
\includegraphics[height=0.75in]{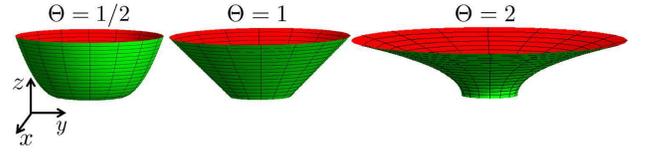}
\caption{\label{fig:surfaces}  (Color online) Three examples of surfaces generated by the set of points $\mathbf{r}(t,\phi)$ given by Eq.~(\ref{eq:surfparam}) with $0 \leq t \leq 2$ and $-\pi \leq \phi <  \pi$.  We set the initial population circumference and the crossover time to one: $R_0= t^* = 1$.  The $\Theta = 1/2$ surface is shaped like a bowl and has positive Gaussian curvature $K>0$. Conversely, the $\Theta = 2$ surface is shaped like a trumpet and has  $K<0$.  Confining a population to grow along the $\Theta = 1$ surface (with $K = 0$)  is another way of implementing the regular radial range expansion with linear inflation.     }
\end{figure}
The  $\Theta \neq 1$ case might be used to model frontier expansions on curved surfaces, such as a parabolic surface with positive Gaussian curvature, or a negatively curved surface like the horn of a trumpet as shown in Fig.~\ref{fig:surfaces}.  Each position $\mathbf{r}=(x,y,z)^T$ on such a surface can be parameterized by the time $t$ and an angle $\phi \in [-\pi,\pi)$.  The angle $\phi$ parameterizes a location on an expanding circumference.  In Cartesian coordinates,
\begin{equation}
\mathbf{r} (t,\phi)= \begin{pmatrix} R_0 \left[1+ \left( \frac{t}{t^*} \right)^{\Theta} \right] \cos \phi \\ R_0 \left[1+ \left( \frac{t}{t^*} \right)^{\Theta} \right] \sin \phi \\ t  \end{pmatrix} \label{eq:surfparam}.
\end{equation}
The population grows upward along the positive $z$-axis, so that we can identify this direction with the time $t$. The Gaussian curvature $K(t, \phi)$ at each point $(t,\phi)$ of  such a surface can be calculated using standard methods \cite{diffgeom}:
\begin{equation}
K(t,\phi)=\frac{t^2  \left(\frac{t}{t^*}\right)^{\Theta }  \Theta(1-\Theta ) }{\left[1+\left(\frac{t}{t^*}\right)^{\Theta }\right] \left[t^2+R_0^2 \left(\frac{t}{t^*}\right)^{2 \Theta } \Theta ^2 \right]^2}.
\end{equation}
Note that $K(t,\phi)$ is positive when $\Theta < 1$ and negative when $\Theta > 1$.  Surfaces at three different values of $\Theta$ are shown in Fig.~\ref{fig:surfaces}.

The analysis will be general for now, and we will consider circular population fronts growing with arbitrary power  $\Theta > 0$ of $t$, with radius $R(t)$ given by Eq.~(\ref{eq:growingradius}).  A location along the population front is  given by  the $x$ and $y$ components of $\mathbf{r}$ in Eq.~(\ref{eq:surfparam}):  $R(t)\{\cos \phi, \sin \phi\}$, where $\phi \in (-\pi,\pi]$ is the angle parameterizing a location on an expanding circumference. Setting  $x = R(t) \phi$  in Eqs.~(\ref{eq:LangevinVMfull}) and (\ref{eq:LangevinCorrVMfull}), we obtain (upon setting $\mu_b=0$) the Langevin equation for a radial expansion
\begin{align}
  \partial_t f  (\phi,t) &= \dfrac{ D}{R(t)^2} \, \partial^2_{\phi} f(\phi,t)+sf(\phi,t)[1- f(\phi,t)] \nonumber \\
& \qquad  \mbox{}-f(\phi,t)\mu_f+  \eta(\phi,t) 
\label{eq:RadLangevin} ,
\end{align}
where  the noise $\eta(\phi,t)$ obeys
\begin{equation}
\left\langle\eta(\phi,t)\eta(\phi',t) \right\rangle  =   \dfrac{2  af(\phi,t)[1-f(\phi,t)]}{ \,R(t)} \, \delta(t-t') \delta(\phi-\phi'),\label{eq:RadLangevinnoise}
\end{equation}
and we have set the generation time to one: $\tau_g = 1$.  When we compare our simulation results to these voter models for radial range expansions,  we will use an effective lattice constant $a_r$  and coordination number $z_r$, as discussed in Sec.~\ref{SSimulations}.  The corresponding diffusion constant is then  $D=D_r \approx a_r^2/(z_r  \tau_g)$, with $\tau_g = 1$.   Similar equations to Eq.~(\ref{eq:RadLangevin}) and Eq.~(\ref{eq:RadLangevinnoise}) were derived via the stepping stone model in Ref.~\cite{KorolevBac,KorolevRMP} for the linearly inflating case, where $R(t) = R_0 + vt$.

  \subsection{\label{SSNeutralEvo}Neutral Evolution}

An important exactly soluble case is a {\it neutral} radial range expansion,  considered previously  for   linear inflation ($\Theta = 1$ in Eq.~(\ref{eq:growingradius})) and inflation with a rough front \cite{KorolevRMP, KorolevBac, nelsonhallatschek, roughconformal}. We consider first the dynamics without mutations for simplicity.  An illustration of a range expansion undergoing neutral dynamics for a well mixed initial homeland with a fraction $f_0=1/2$ of green (light gray in print) cells is given in the left panels of  Fig.~\ref{fig:structure} $(a)$ and $(b)$.  We see in the figure that the domain walls between green and red (light and dark gray, respectively, in print) sectors will initially annihilate in pairs due to genetic drift leading to domain coarsening.  A useful measurement of the coarsening is the angular heterozygosity (for a spatially homogeneous initial condition)
\begin{align}
H(\phi,t) & \equiv  \left\langle \vphantom{\frac{1}{2}} f(\phi+\phi',t)[1-f(\phi',t)] \right. \nonumber \\
&  \qquad \qquad  \left. \vphantom{\frac{1}{2}} {} + f(\phi',t)[1-f(\phi+\phi',t)]\right\rangle_{\phi'},
\end{align}
where  $\langle \quad \rangle_{\phi'}$ includes an angular average over $\phi'$ and over  realizations of the noise in the Langevin equation (Eq.~(\ref{eq:RadLangevin})) for $f(\phi,t)$.  The heterozygosity $H(\phi,t)$ represents the probability of observing two cells of \textit{different} types an angular distance $\phi$ apart at time $t$.  Since our voter-type model has a single cell at each lattice point, $H(\phi,t)$ satisfies the boundary condition  $H(\phi=0,t) = 0$.

We set $s =\mu_f= 0$ in Eq.~(\ref{eq:RadLangevin}) and use the \^Ito calculus to get the  equation obeyed by the angular heterozygosity $H(\phi,t) $ for the initial condition $f(\mathbf{x},t=0) \equiv f_0$.  We also assume that we can extend the range of $\phi$ from the interval $(-\pi,\pi]$ to the entire real line $(-\infty,\infty)$.  This approximation will be valid as long as the average angular sector size is small compared to $2 \pi$, the angular size of the population  \cite{nelsonhallatschek}.  The heterozygosity $H(\phi,t)$ for an arbitrary power law inflation $R(t) = R_0[1+(t/t^*)^{\Theta}]$ obeys a  diffusion equation with a time-dependent diffusivity $D(t)=2 D_r/[R(t)]^2$. The time dependence can be removed by introducing  a rescaled, ``conformal'' time variable
\begin{align}
t_c(\Theta,t) & \equiv  \frac{t}{\Theta} \left\{   \frac{1}{1+(t/t^*)^{\Theta}}\right. \nonumber  \\
&  \left. {}+(\Theta-1) {}_2F_1\left[1,\frac{1}{\Theta}
; 1+ \frac{1}{\Theta}; - \left( \frac{t}{t^*} \right)^{\Theta} \right]  \right\} , \label{eq:conformaltime}
\end{align}
 where ${}_2F_1(\alpha,\beta; \gamma;z)$ is the Gauss hypergeometric function \cite{ryzhik}.  Changing variables to the conformal time yields a simple diffusion equation for $H(\phi,t_c)$, namely
\begin{equation}
\partial_{t_c} H(\phi,t_c) = 2 D_r \, \partial_{\phi}^2 H(\phi,t_c). \label{eq:hetsimplediff}
\end{equation}
Eq.~(\ref{eq:hetsimplediff}) can now be solved for the uniform initial condition $H(\phi,t_c=0) = H_0 \equiv 2 f_0(1-f_0)$ and  the boundary condition $H(\phi=0,t_c)=0$.  The solution, after changing back to the original time variable, is
\begin{align}
H(\phi,t) 
& = H_0 \, \mbox{erf} \left[ \frac{R_0|\phi|}{ \sqrt{ 8D_r t_c (\Theta,t)}} \right] , \label{eq:heterorad}
\end{align}  
where $t_c(\Theta,t)$ is given by Eq.~(\ref{eq:conformaltime}).

We recover the known solution for a linear range expansion by replacing $t_c(\Theta,t)$ with $t$ in   Eq.~(\ref{eq:heterorad})  \cite{KorolevRMP}.   The limiting value $t_c(\Theta,t \rightarrow \infty)$ is particularly interesting as it tells us for which values of $\Theta$ we can expect a ``cutoff'' in the dynamics, i.e., when the time variable $t_c(\Theta,t)$ approaches a finite value and $H(\phi,t)$ approaches a limiting shape as $t \rightarrow \infty$ in the radial expansion.  In particular, we find
\begin{equation}
t_c(\Theta,t\rightarrow\infty) = \begin{cases} 
\infty & \Theta \leq \frac{ 1}{2} \\[8pt]
\dfrac{\pi (\Theta-1)t^*}{ \Theta^2 \sin(\pi/\Theta)}& \Theta > \frac{1}{2}
\end{cases}.
\end{equation}
Thus, when the  exponent $\Theta$ is larger than the characteristic value 1/2 (which matches the diffusive wandering behavior of domain walls), the inflation will eventually dominate the diffusive dynamics.

The inflation-induced crossover also modifies  the average angular sector width $\Delta \phi(t)$. The width follows from Eq.~(\ref{eq:heterorad}), upon noting that   (see Eq.~(C2) in Ref.~\cite{KorolevRMP})
\begin{equation}
\Delta \phi (t)= \left. \left( \frac{\partial H}{\partial \phi} \right)^{-1} \right|_{\phi \rightarrow 0^+} = \frac{\sqrt{2 \pi D_r t_c(t,\Theta)}}{R_0 H_0} 
.  \label{eq:angularwidth}
\end{equation}
Eq.~(\ref{eq:angularwidth}) will be valid only when $\Delta \phi(t) \ll 2 \pi$ since we do not take into account sectors that take over the entire population.  When $\Theta \leq 1/2$, the angular width $\Delta \phi(t)$ diverges at long times  so that only a \textit{single} genetic variant is present.   Note the striking difference between this behavior and the $\Theta > 1/2$ case where the number of surviving sectors approaches a finite value given by
\begin{equation}
N_{\mathrm{surv}} = \frac{2 \pi}{\lim_{t \rightarrow \infty} \Delta \phi(t)} =  R_0 H_0 \sqrt{ \frac{2\Theta^2 \sin(\pi/\Theta)}{  D_r (\Theta-1)t^*}}. \label{eq:secnum}
\end{equation} 
This result is in agreement with the results of Refs.~\cite{KorolevRMP, nelsonhallatschek} for the case  $\Theta = 1$ and $t^* = R_0/v$.

 \begin{figure}
\includegraphics[height=2.5in]{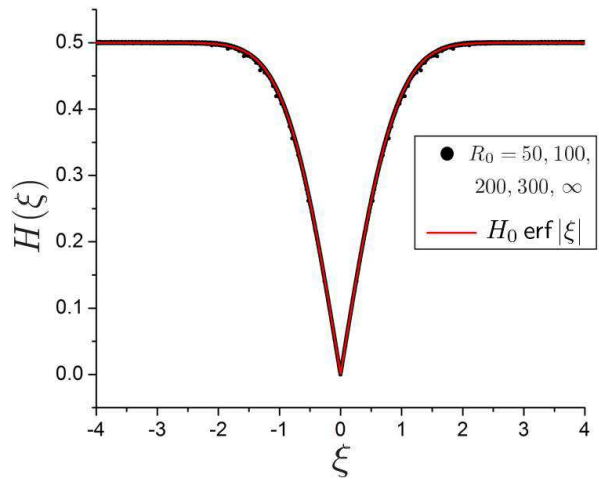}
\caption{\label{fig:hetcollapse}  (Color
  online) Nearly perfect collapse of the simulation data using the scaled variable $\xi = R_0 \phi/\sqrt{8 D_r t_c(t)}$ for the heterozygosity with $H_0=1/2$, corresponding to random initial conditions, with equal probabilities for the two different colors. The initial colony radius is given by $R_0=50,100,200,300$. Linear range expansions correspond to the limit $R_0 \rightarrow \infty$.   All of the radial expansion simulations were fit with a single parameter $D_r$. The scaling variable is $\xi = x/\sqrt{8 D t}$ for linear expansions. }
\end{figure}

We now specialize to $\Theta = 1$ and $t^* = R_0/v$ in order to check our results with simulations and known results.  In this case, we have the simple relation
\begin{equation}
t_c(t) \equiv t_c(1,t)  = \frac{t}{1+t/t^*}  =\frac{ t}{1 + vt/R_0}. \label{eq:mobiustransform}
\end{equation}
 The transformation from $t$ to $t_c(t)$ in Eq.~(\ref{eq:mobiustransform}) is a special case of the Mobi\"us transformation and maps the interval $t \in (0,\infty)$ to $t_c \in (0, t^*)$ \cite{complex}. Both the linear and radial expansion heterozygosities collapse to a single universal function $H(\xi)= H_0\, \mbox{erf} \, |\xi|$, with $\xi=x/\sqrt{8 D t}$ for linear expansions, and $\xi = R_0 \phi/\sqrt{8 D_r t_c(t)}$ for radial ones.   The comparison requires fitting the radial diffusion constant $D_r = a_r^2/(z_r \tau_g)$, which depends on the effective radial lattice constant $a_r$ and the coordination number $z_r$ for the Bennett model cellular packing ($\tau_g=1$ in the simulations).  We find excellent data collapse and agreement between simulation and the voter model theory.  These theoretical predictions for the heterozygosity are described in more detail in Ref.~\cite{KorolevRMP} and  are compared to simulation data in Fig.~\ref{fig:hetcollapse}.  The connection between the radial and linear range expansions is also discussed in Ref.~\cite{roughconformal} for expansions with rough fronts.

\subsection{\label{SSMutations} Heterozygosity in Radial Range Expansions}

 It is also possible to derive analytic solutions for the heterozygosity in the biologically relevant limit of neutral mutations.  Neutral mutations are common and can have an important effect on a population's genetic composition  \cite{kimura}.  We now compare strictly neutral mutational dynamics with genetic drift in radial versus linear range expansions.  Many previous results for the linear case are reviewed in Ref.~\cite{KorolevRMP}.

Mutations spoil the clean connection between  linear and radial range expansions via a conformal time coordinate (Eq.~(\ref{eq:mobiustransform})).  Instead, mutations (we now assume both $\mu_f$ and $\mu_b$ are nonzero)  dominate the dynamics at long times in both types of range expansion. The equation of motion for $H(\phi,t)$ can be derived from Eq.~(\ref{eq:RadLangevin}) via the \^Ito calculus, just as in the neutral case.  An additional term $\mu_b [1-f(\mathbf{x},t)]$ is added to the left-hand side of Eq.~(\ref{eq:RadLangevin}) to take into account backward mutations from red (dark gray) to green (light gray) cells at rate $\mu_b$.  The equation of motion for a radial range expansion for a homogeneous initial cell density $f(\mathbf{x},t=0)=f_0$  is  
\begin{align}
\partial_t H(\phi,t) &   = \frac{2  D_r}{[R(t)]^2} \partial_{\phi}^2 H(\phi,t)-2(\mu_{f}+\mu_{b})H (\phi,t) \nonumber \\
& \quad {} + 2(\mu_{f}-\mu_{b})f(t)+2 \mu_b, \label{eq:radhetmut}
\end{align}
where the third and fourth terms on the right-hand side of Eq.~\ref{eq:radhetmut} come from the coupling between the equations for the first and second moments of   $f(\mathbf{x},t)$. These terms include a contribution from the average fraction of green cells $f(t) \equiv \langle f(\mathbf{x},t) \rangle_{\mathbf{x}}$, given by
\begin{equation}
f(t) =  \frac{\mu_b+ [f_0  (\mu_f+\mu_b)-\mu_b]e^{-t (\mu_f +\mu_b )}   }{\mu_f+\mu_b },
\end{equation}
(see Ref.~\cite{KorolevRMP}) which decays to a steady state value $f(t \rightarrow \infty)=\mu_b/(\mu_f+\mu_b)$. To get the analogous linear equation,  one can replace the angular coordinate $\phi$ with a spatial coordinate $x$ along the linear front and set $R(t) \rightarrow 1$.   

The solution to Eq.~(\ref{eq:radhetmut}) for an initially homogeneous random initial condition $H(\phi,t=0)=H_0 =2 f_0(1-f_0)$ is
\begin{align}
H(\phi,t) = H_{\infty}[(R_0+vt)\phi]+ e^{-(\mu_{f}+ \mu_{b})t} H_{\mathrm{\,tr}}(\phi,t), \label{eq:hetmutfull}
\end{align}
which at long times approaches the steady state $H_{\infty}[(R_0+vt) \phi]$, where  
\begin{equation}
H_{\infty}(x)=\frac{2\mu_{f} \mu_{b} }{(\mu_{f}+ \mu_{b})^2} \left[1 - e^{-|x| \sqrt{ \frac{\mu_{f}+ \mu_{b}}{D_r}}} \right].
\end{equation}
The transient term $H_{\mathrm{tr}}(\phi,t)$ is given by
\begin{align}
&H_{\mathrm{tr}}(\phi, t)=H_0  e^{-t'}  \, \mbox{erf} \,\left[ \frac{ \xi_r}{\sqrt{1+t/t^*}}\right] \nonumber \\
& +\frac{2t'( \mu_f-\mu_b)[f_0 \mu_f-(1-f_0)\mu_b] e^{-t'} }{ (\mu_f + \mu_b)^2}  \nonumber \\ & \qquad \qquad \quad \, \, \, \,  \times\int_0^{1} \mbox{erf} \left[\frac{   \sqrt{t^*/t+y} \, \xi_r }{ \sqrt{ (1+t^*/t) (1-y)}}\right]  e^{  t' y }  \mathrm{d} y \nonumber  \\
& +\frac{4t'  \mu_f\mu_b e^{-t'} }{ (\mu_f + \mu_b)^2 } \int_0^{1} \mbox{erf} \left[ \frac{   \sqrt{t^*/t+y} \, \xi_r }{ \sqrt{ (1+t^*/t) (1-y)}}\right]  e^{ 2t' y }  \, \mathrm{d} y  \nonumber \\
& + \frac{2e^{ t'} \mu_f \mu_b}{(\mu_f+\mu_b)^2}  \left[ e^{-2 \sqrt{2t'} \,\xi_r}-1\right], \label{eq:Hradtrans}
\end{align}
where we have defined a dimensionless time $t' \equiv (\mu_f+\mu_b) t$, and a dimensionless length scale
\begin{equation}
\xi_r =  \frac{(R_0+vt) \phi}{\sqrt{8 D_r t}}.
\end{equation}
The steady-state shape $H_{\infty}[(R_0+vt) \phi]$ is exactly the same for linear range expansions  with mutations (see Ref.~\cite{KorolevRMP}) if we identify $(R_0 + vt) \phi = x$ as the separation between two cells in the linear expansion.  
In both expansions, mutations set the average domain size at long times. Also, a careful asymptotic analysis of the integrals in Eq.~(\ref{eq:Hradtrans}) reveals that the transient term in Eq.~(\ref{eq:hetmutfull}) decays as $e^{-(\mu_f+\mu_b)t}$.  A similar analysis shows that Eq.~(\ref{eq:Hradtrans}) reduces to the correct transient term for a linear range expansion in the limit $t^* \rightarrow \infty$.  More details about the linear range expansion heterozygosity are discussed in Ref.~\cite{KorolevRMP}.

Eq.~(\ref{eq:Hradtrans}) reveals two competing time scales:  the mutational relaxation time $(\mu_f+\mu_b)^{-1}$ and the inflation time $t^* = R_0/v$.  Mutations equilibrate on the first time scale, while the second determines when inflation sets in. The effect of inflation is observable for intermediate time scales such that $t^* \ll t \ll (\mu_f+\mu_b)^{-1}$.  The average angular domain size $\Delta \phi (t)= [ \left. \partial_{\phi} H(\phi,t) \right|_{\phi=0}]^{-1}$ in this intermediate time regime is given by
\begin{align}
\Delta \phi (t) & =  \frac{ \sqrt{2 \pi  D_r t_c(t)}}{H_0R_0} \Bigg\{1  -2t' \nonumber \\
& + \left[ \frac{\pi}{H_0} \left[ f_0+ \frac{\mu_b(1-2f_0)}{\mu_b+\mu_f}\right] \sqrt{\frac{t}{t^*}} + \mathcal{O} \left( \sqrt{ \frac{t^*}{t}} \right)  \right]t' \nonumber \\
& \qquad \qquad \qquad \qquad  \qquad \qquad \quad     + \mathcal{O}(t'^2) \Bigg\}^{-1} , \label{eq:mutangularwidth}
\end{align}    
where $t' = (\mu_f+\mu_b)t \ll 1$ and $t_c(t)$ is the conformal time coordinate (Eq.~(\ref{eq:mobiustransform})). The factor in front of the curly braces in Eq.~(\ref{eq:mutangularwidth}) is the neutral result of Eq.~(\ref{eq:angularwidth}) with $\Theta = 1$.

\begin{figure}
\includegraphics[height=2.3in]{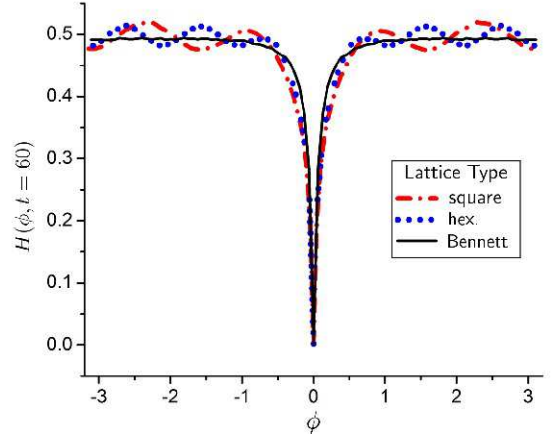}
\caption{\label{fig:latticedefects} (Color
  online) A calculation of the heterozygosity $H(\phi,t)$ at time $t = 48 \gg t^*$ on different lattices. $R_0 = 16$, $v  = 1$, $\mu_f =0.1$, and $s = 0.57$ in the simulations.  The length and time units are measured using the rescaling discussed in Sec.~\ref{SSimulations}.  Oscillations in the heterozygosity on regular lattices reflect the $4-$ and $6-$fold  rotational symmetries.  These oscillations are absent for the amorphous Bennett model lattice with $\rho = 0.769$ and $Q = 0.5$.}
\end{figure}

The heterozygosity  correlations are sensitive to dynamics of our model on different lattices.  When we are in the active phase of the dynamics  with selection and one-way mutations $(\mu_b=0)$ on a regular lattice, the heterozygosity function exhibits  oscillations as shown in Fig.~\ref{fig:latticedefects}.  The number of oscillations reflects the rotational symmetry of the lattice, as shown for  the hexagonal (6 oscillations) and square (4 oscillations) lattices in Fig.~\ref{fig:latticedefects}.  The oscillations arise because the green cell domains grow preferentially along the crystallographic directions of the lattice. These lattice artifacts obscure the evolutionary dynamics. As discussed in Sec.~\ref{SSimulations}, the problem is corrected by using an amorphous lattice which smooths out the oscillations.

\subsection{\label{SSOthers} Single Seed Dynamics and Selective Advantage}

\begin{figure}
\includegraphics[height=1.5in]{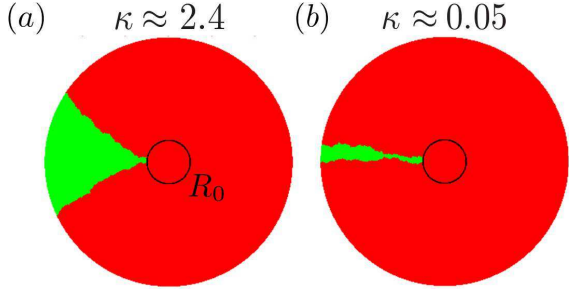}
\caption{\label{fig:SingleSeed} (Color online) Two examples of single sector dynamics with $R_0 = 50$ (in units of the average cell diameter) evolved over about 250 generations for two very different values of the key inflationary parameter $\kappa \approx s \sqrt{R_0/2v \tau_g} \,$ discussed in the text.  In $(a)$ a combination of selection bias and inflation dominates genetic drift after a short initial transient, with $\kappa \approx 2.4$. The initial sector width in radians at the edge of the homeland  is $\phi_0 \approx 0.1$.  In $(b)$ $\kappa \approx 0.05 \ll 1$, so genetic drift can influence the result expected from selection bias and inflation alone for an extended period.  We set $\phi_0 \approx 0.2$ in this case.  We see that the fate of the sector in $(a)$ is decided much faster than the fate of the sector in $(b)$, which still might eventually die out. }
\end{figure}

Finding  exact solutions is difficult when we include selection.  Unlike the cases considered in the  two previous sections, it is not possible to apply the \^Ito calculus to Eq.~(\ref{eq:RadLangevin}) to exactly calculate the angular heterozygosity $H(\phi,t)$.  The nonlinear term $s f(\mathbf{x},t)[1-f(\mathbf{x},t)]$ in Eq.~(\ref{eq:RadLangevin}) leads to  an infinite  set of equations involving ever higher order moments of the field $f(\mathbf{x},t)$.  Although various approximations can be tried, accurate analytic results are limited \cite{KorolevRMP}.    Hence, to study the effects of selection on radial range expansions  we instead consider the fate of a ``seed,'' i.e. a single initial green sector of size $\phi_0 \ll 1$ with selective advantage $s$. We assume that mutations are rare, so that  $\mu_f = \mu_b \approx 0$.   During each generation (corresponding to a time step $\tau_g$), the size $\phi$  only changes due to the competition between red and green cells at the two boundaries of the sector. We make the further simplifying assumption, easily imposed in our simulations, of neglecting the bulge expected for a seed representing a favorable mutation at the frontier \cite{nelsonhallatschek,KorolevMueller}.  This approximation should be adequate for $0 < s \ll 1$, and it allows analytic progress for nearly circular colonies.

Each generation in the radial expansion simulations consists of a rim of cells with an approximately single cell thickness (see Sec.~\ref{SSimulations}).  Hence, the green sector size  changes  each generation due to a single green cell competing with a neighboring red cell at each of two boundaries.  Specifically, each boundary can move by some average amount $\pm \Delta\phi \approx  \pm  \tilde{a}_r/R(t)$ per generation, where $\tilde{a}_r$ is an effective spacing between adjacent cells along the circular population front  approximately equal to the effective lattice spacing $a_r$ (see Appendix~\ref{App:SingleSectorInfl}).  The probability distribution for the angular size $\phi$ of  a sector  satisfies the Fokker-Planck-like equation
(also known as a Kolmogorov forward equation -- see Ref.~\cite{KorolevRMP})\begin{equation}
\partial_{\tau} p_s(\phi, \tau)  =- \frac{w}{1-\tau} \, \frac{\partial p_s}{\partial \phi} + \Delta\, \frac{\partial^2 p_s}{\partial \phi^2}, \label{eq:ssFPE}
\end{equation}
where $\tau = t_c(t)/t^* = vt/(R_0+vt)=vt/R(t)$ is a dimensionless conformal time coordinate.   The term in Eq.~(\ref{eq:ssFPE})  proportional to $w$ represents  a bias due to selection, and the term proportional to $\Delta$ represents genetic drift.  As shown in Appendix~\ref{App:SingleSectorInfl}, $w$ is proportional to the selective advantage enjoyed by the green cells, 
 \begin{equation}
w  \approx\gamma\,\frac{\tilde{a}_rs}{v \tau_g}  \qquad (\mbox{for } s \ll 1), \label{eq:wdef}
\end{equation}
and
\begin{equation}
\Delta   \approx \frac{ \tilde{a}_r^2}{R_0 v \tau_g} \qquad (s \ll 1) , \label{eq:nudeltadefs}
\end{equation} 
where $\gamma$ is a lattice-dependent factor of order unity.  For the disordered Bennett model lattice used in the rest of the paper (with $Q=0.6$ and $\rho=8/11$), we find $\tilde{a}_r \approx 0.90$ (in units of the average cell diameter) and $\gamma \approx 1.2$.    We impose the absorbing boundary condition $p_s(\phi=0,\tau) = 0$, just as in the linear case \cite{nelsonhallatschek}.  Note from Eq.~(\ref{eq:ssFPE}) that inflation suppresses both drift and diffusion by folding the entire time evolution of the range expansion into the conformal time $\tau$. Inflation  suppresses the diffusion  more than the drift term since they scale with $R(t)^{-2}$ and $R(t)^{-1}$, respectively.

 The  Langevin equation corresponding to Eq.~(\ref{eq:ssFPE}) is 
\begin{equation}
\frac{d \phi}{d \tau} = \frac{w}{1-\tau}+ \sqrt{2 \Delta} \, \eta(t), \label{eq:lansec}
\end{equation}
where $\eta(t) $ is a Gaussian white noise (with $\langle \eta(t) \eta(t') \rangle = \delta(t-t')$).  Note that unlike the multiplicative noise in the Langevin equation for the coarse-grained cell density $f(\mathbf{x},t)$ (Eq.~(\ref{eq:LangevinVMfull})),  Eq.~(\ref{eq:lansec}) has a simple noise term. 
 Hence, single sector dynamics is more easily analyzed using the ``dual'' description of sector boundary motion, instead of  the full density $f(\mathbf{x},t)$ dynamics in Eq.~(\ref{eq:LangevinVMfull}) (also see Appendix~\ref{App:SingleSectorInfl}).  For large sector sizes $\phi \gg \tilde{a}_r/R_0$, we might hope to neglect genetic drift relative to the bias term.  Upon setting $ \Delta = 0$, Eq.~(\ref{eq:lansec}) is solved by
\begin{align}
\phi(\tau) & = \phi_0 - w  \ln \left(1-\tau\right)  = \phi_0 + w \ln \left[ \frac{R(\tau)}{R_0} \right], \label{eq:determsector}
\end{align}
similar to the logarithmic spiral sector boundaries found in Ref.~\cite{KorolevMueller}.  To obtain the conformal time to fixation $\tau_f$, we  set $\phi(\tau_f) = 2 \pi$ to find
\begin{equation}
 \tau_f = 1-\exp \left[ -\frac{2\pi- \phi_0}{w } \right]
\end{equation}
for the purely deterministic model.  When noise is included in our radial voter model, we find from our simulations that $\tau_f \approx 1$ for all $s \lesssim 0.5$ and $\phi_0 \lesssim \pi$.  Thus, a sector with small initial angular width $\phi_0 \ll 1$ can only take over a population in a reasonable time provided $w \propto s$ is large. When $\phi_0$ is very small, the takeover predicted by the deterministic solution in Eq.~(\ref{eq:determsector}) is preempted by number fluctuations, i.e., genetic drift. If a small initial sector survives the genetic drift, it can eventually take over the entire population in a deterministic fashion.  
\begin{figure}
\includegraphics[height=2.1in]{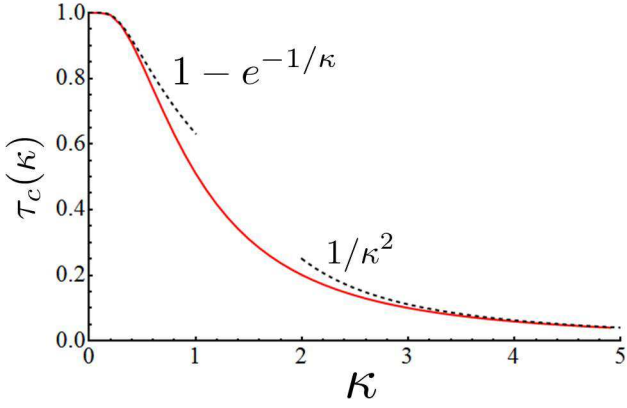}
\caption{\label{fig:tauc} (Color online) A plot of the crossover time $\tau_c$ that marks when genetic drift is no longer able to keep up with inflation.  The time  $\tau_c$ is  given as a dimensionless time coordinate,  related to the  time $t$ via  $\tau =  vt/(R_0+vt)=vt/R(t)$.  The dashed lines show the asymptotic approximations to $\tau_c$ given in Eq.~(\ref{eq:taucasmyp}).               }
\end{figure}

To go beyond the deterministic approximation, we must account for the diffusive dynamics of the sector boundaries caused by genetic drift.  To treat the case $\Delta > 0$, we define a new angle variable
\begin{equation}
\psi(\tau) = \phi(\tau)+  w \ln (1-\tau),
\end{equation}
such that the sector probability distribution $p_s(\psi,t)$ now obeys
\begin{equation}
\partial_{\tau} p_s(\psi, \tau) =   \Delta \, \frac{\partial^2 p_s(\psi,\tau)}{\partial \psi^2} \label{eq:diffvarphi}.
\end{equation}
Our boundary condition $p_s(\phi=0, \tau) = 0$ moves in time, $p_s(\psi = w \ln(1-\tau),\tau) = 0$.  The diffusive motion of the variable $\psi$, with uncertainty $\delta \psi = \pm \sqrt{2 \Delta\tau }$, will be unable to keep up with the absorbing boundary condition at $\psi = w \ln(1-\tau)$ after a crossover time $\tau_c$, such that
\begin{equation}
\sqrt{\tau_c} =  -  \frac{w}{\sqrt{2 \Delta}} \, \ln (1 -\tau_c) \equiv  - \kappa \ln (1 - \tau_c), \label{eq:sscrittime}
\end{equation}
with the key parameter
\begin{equation}
\kappa \equiv \frac{w}{\sqrt{2 \Delta}}\approx s \sqrt{\frac{ t^*}{ 2 \tau_g }} \,\approx s \sqrt{\frac{R_0}{2v \tau_g}} .
\end{equation}
A plot of $\tau_c(\kappa)$ is shown in Fig.~\ref{fig:tauc}. The behavior in the two limits $\kappa \gg 1$ and $\kappa \ll 1$ is given by
\begin{equation} \label{eq:taucasmyp}
\tau_c(\kappa) \approx \begin{cases}
1- e^{-1/\kappa }& \kappa \ll 1 \\
\kappa^{-2} & \kappa \gg 1
\end{cases}.
\end{equation}

When $\kappa \gg 1$ (strong selection), the transition from trajectories dominated by genetic drift to those dominated by deterministic natural selection is very fast. The inflationary effects on the survival probability are then negligible, and the relevant time scale is the diffusion time $\tau_l \approx \phi_0^2/(2 \Delta)$.  For times $\tau \gg \tau_l$, the survival probability of a sector for radial voter models is  described by the long-time linear range expansion result \cite{nelsonhallatschek}
\begin{equation}
S_{r}(\tau \gg \tau_l) \approx  \frac{1- e^{-w \phi_0/ \Delta }}{1- e^{-2  \pi w / \Delta}}  \qquad (\kappa \gg 1).
\end{equation}
If $\tau \ll \tau_l$, however, we must either use the full linear survival probability solution of Eq.~(\ref{eq:ssFPE}) (Eq. 3.2.15 of Ref. \cite{redner}), or else exploit the small time solution 
\begin{equation} 
S_r(\tau \ll \tau_l) \approx 1- \frac{\left(e^{-w \phi_0/\Delta}+1 \right)}{2}\, \mbox{erfc}\left[ \frac{\phi_0}{\sqrt{4 \Delta \tau}} \right] \quad (\kappa \gg 1). \label{eq:linsurvpsmalltime} 
\end{equation} 
    In our simulations, it is difficult to access the regime $\kappa \gg 1$  for small $s$ since this requires  large values of the initial radius $R_0$.   In practice, we are limited to the range $0 \leq \kappa \lesssim 10$.  Fig.~\ref{fig:SingleSeed}$(a)$ shows a simulation for $\kappa \approx 2.4$, with only minor diffusion motion in the sector boundaries visible at the onset of the range expansion.  In this case, if the sector survives beyond the relatively small time $\tau_c$,   it is quite likely to survive indefinitely.

\begin{figure}
\includegraphics[height=2in]{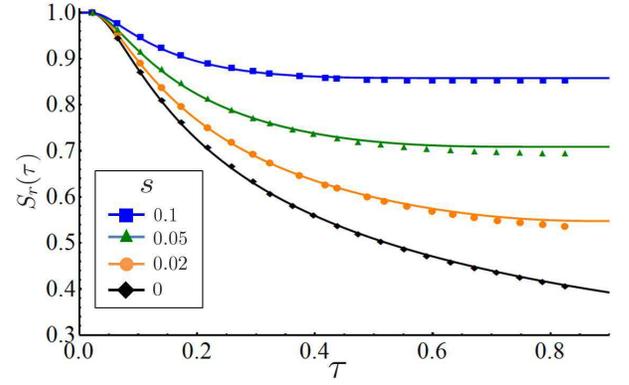}
\caption{\label{fig:radsurvp} (Color online) Radial survival probabilities  $S_r(\tau)$ for a green sector (like that in Fig.~\ref{fig:SingleSeed}) of size $\phi_0 \approx 0.036$ initialized on the rim of a homeland with $R_0 = 300$, evolved over 1750 generations, for a variety of selective advantages $s$.    The solid lines represent Eq.~(\ref{eq:radsurvint}), the adiabatic approximation to $S_r(\tau)$.     The integral in Eq.~(\ref{eq:radsurvint}) is evaluated numerically.  The  survival probabilities were averaged over  $8 \times 10^4$ independent runs. We fit the black line (the $s=0$ case) with the lattice parameter $\tilde{a}_r$, finding $\tilde{a}_r \approx 0.9$.  The other lines use $\gamma \approx 1.2$ (see Appendix~\ref{App:SingleSectorInfl}).       }
\end{figure}

  When  $\kappa \ll 1$,  however, genetic drift dominates the deterministic sector size prediction in Eq.~(\ref{eq:determsector}) over the entire interval $\tau \in (0, \tau_c)$.  As discussed above, a single sector is quite unlikely to take over the whole population for   values of $\phi_0$ not too close to $2 \pi$.   We can then  approximate the diffusive motion of $\psi = \phi+w \ln(1-\tau)$ by confining it to a semi-infinite interval  between $b(\tau) \equiv w\ln(1-\tau)$ and $\infty$, instead of imposing an upper limit of $2 \pi$.

When $\tau < \tau_c$, the boundary condition $p_s(\psi = w \ln(1-\tau),\tau) = 0$ moves slowly compared to the diffusion, allowing an adiabatic approximation \cite{redner}: The Fokker-Planck equation (Eq.~(\ref{eq:ssFPE}))  for the sector size probability distribution $p_s(\phi,\tau)$  is treated as a drift-diffusion equation with diffusion coefficient $\Delta$ and a time-dependent drift velocity   $|d b(\tau)/d \tau|=w/(1-\tau)$, evaluated instantaneously at $\tau$.  With the initial probability distribution $p_s(\phi,0) = \delta(\phi-\phi_{0})$, the solution to Eq.~(\ref{eq:ssFPE}) in this approximation follows from the method of images \cite{redner}:
\begin{align}
p_s(\phi, \tau) & \approx\frac{1}{\sqrt{4 \pi  \Delta \tau}} \left[e^{-\frac{\left(\phi -\phi_{ 0}-w\tau/(1-\tau)\right)^2}{4 \Delta \tau }}   \right. \nonumber \\
& \qquad \qquad  -\left. e^{ -\frac{w \phi_0}{\Delta (1-\tau) }}e^{-\frac{(\phi +\phi_{0}-w \tau/(1-\tau))^2}{4 \Delta \tau }} \right] . \label{eq:Psecrad}
\end{align}
The change $dS_r(\tau)/d \tau$ in the radial survival probability  must be given by the probability flux $\left. [\Delta \partial_{\phi} p_s(\phi,\tau)] \right|_{\phi=0}$ evaluated at the absorbing boundary.  Integrating this probability flux for a radial expansion yields 
\begin{align}
S_r(\tau) & =1-\Delta \int_{0}^{\tau} \mathrm{d} \tau' \, \left. \left[ \partial_{\phi} p_s(\phi,\tau')  \right] \right|_{\phi=0}\nonumber \\[5pt]
 & \approx1-  \int_{0}^{\tau} \mathrm{d} \tau' \, \frac{\phi_0e^{-\frac{(\phi_0(1-\tau')+w\tau')^2}{4 \Delta (1-\tau')^2 \tau'}}}{\sqrt{4 \pi \Delta (\tau')^3 }} \label{eq:radsurvint} \\
& \approx 1- e^{- \frac{\phi_0 w}{2\Delta}}\left[1+ \frac{w \phi_0^2 }{4 \Delta^2 } \left( \phi_0 + \frac{ w}{2 }\right) \right] \mbox{erfc}\left[\frac{\phi_0}{2 \sqrt{\Delta  \tau }}\right] \nonumber \\
& \quad {}  + \frac{ \sqrt{  \tau}w \phi_0  }{ \sqrt{4 \pi \Delta^3 }  } \left( \phi_0 + \frac{ w}{2 }\right) e^{- \frac{2\phi_0 w \tau+\phi_0^2}{4\Delta \tau}} \quad (\tau \ll 1),  \label{eq:radsurvad}
\end{align}
where an asymptotic (small $\tau$) approximation is used to evaluate the integral in Eq.~(\ref{eq:radsurvint}). Fig.~\ref{fig:radsurvp} shows that Eq.~(\ref{eq:radsurvint})  is a remarkably good approximation to $S_r(\tau)$ even  in the selection dominated regime $\tau > \tau_c$.

 In the weak selection limit  $\kappa \rightarrow 0$, the movement of the boundary condition can be ignored entirely.  Eq.~(\ref{eq:radsurvad}) then reduces to the survival probability of a diffusing particle on a semi-infinite interval with an absorbing boundary at the origin \cite{redner}:
\begin{equation}
S_{r}(\tau)  \,  \stackrel{\kappa \rightarrow 0}{ \longrightarrow} \,  \mbox{erf}\left[\frac{\phi_{0}}{ \sqrt{4\Delta \tau }}\right] . \label{eq:limitingsurvp}
\end{equation}
Fig.~\ref{fig:radsurvp} shows that the theory matches the simulation results for radial expansions with $\phi_0 \approx 0.036$ and $R_0 = 300$ well, especially at small values $\tau=vt/(R_0+vt)$. As $\tau \rightarrow 1$ (i.e. in the limit of long times), both inflation and selection prevent the extinction of the sector and   $S_r(\tau)$  approaches a nonzero limit for all values of $s$, including $s=0$.  An asymptotic analysis of the integral in Eq.~(\ref{eq:radsurvint})  for small $w \propto s$  shows that the limiting value $S_r(\tau \rightarrow 1) \equiv S_{\infty}$  for small initial angular sector size $\phi_0$ is approximately
\begin{align}
&S_{\infty}\approx  \mbox{erf}\left[\frac{\phi_{0}}{ \sqrt{4\Delta  }}\right] + \left[ 1-  \frac{ \ln\left( \frac{w \phi_0}{2 \Delta} \right)}{ \sqrt{4\pi \Delta}}\,\phi_0  \right] \frac{\phi_0w}{2\Delta}  +\mathcal{O}(w \phi_0^2). \label{eq:limitingsurvpwsel}
\end{align}

For completeness, we also consider the regime  $\tau_{d} \equiv 2 \pi^2/\Delta \ll 1$, where genetic drift is strong enough to allow an initially small green sector in Fig.~\ref{fig:SingleSeed} to diffusively take over the entire population.  In this case, the finite range of  the sector size $\phi$ becomes important.  In the absence of selection ($\kappa \rightarrow 0$), the survival probability can be computed using Laplace transform techniques (see e.g., Ref.~\cite{redner,nelsonhallatschek}) and is given by: 
\begin{equation}
S_{r}(\tau)  \,  \stackrel{\kappa \rightarrow 0}{ \longrightarrow} \,   \frac{\phi_0}{2 \pi} + \frac{2}{\pi}\sum_{n=1}^{\infty}  \frac{1}{ n}\, e^{- \Delta n^2 \tau/4} \sin \left[ \dfrac{n \phi_0}{2} \right]. \label{eq:survpfinites}
\end{equation}
Note that the results for the survival probability in Eq.~(\ref{eq:survpfinites}) and Eq.~(\ref{eq:limitingsurvp})  coincide when $\Delta$ is small, and the finite range of the sector size becomes irrelevant.

 \subsection{\label{SSScaling} Directed Percolation With Inflation}

 \begin{figure}
\includegraphics[height=2.3in]{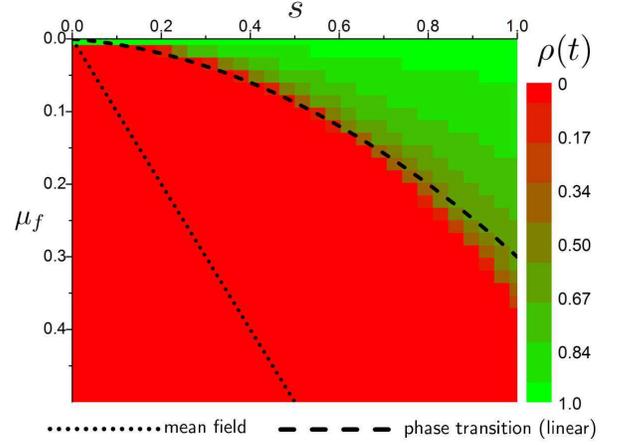}
\caption{\label{fig:DKPhase}  (Color online) The color map of the average fraction of green cells at long times ($t=1800$ generations) for a radial expansion with a homeland radius of $R_0 = 200$, averaged over 400 runs,  for positive values of $\mu_f$ and $s$.  The average fraction $\rho(t)$ approaches a steady state at long times so that $\rho_{\infty} = \lim_{t \rightarrow \infty} \langle f(t) \rangle = \lim_{t \rightarrow \infty} \langle f(\mathbf{x},t)\rangle_{\mathbf{x}} \approx \rho(t=1800)$.  This approximation only breaks down extremely close to the phase boundary between the red (dark gray) and green (light gray) regions and the difference is not visible at the resolution of the plot.  The dotted line shows the mean field transition.  The dashed line shows the approximate position of the directed percolation phase transition for a linear expansion, as shown in Fig.~\ref{fig:PhaseDiagram}. }
\end{figure}

As illustrated in Fig.~\ref{fig:PhaseDiagram}, linear range expansions within the Domany-Kinzel (DK) model on a hexagonal lattice for $\mu_b = 0$ lead to a phase transition driven by competition between the selective advantage $s$ of the green cells and deleterious forward mutations $\mu_f$, from green to red.  In population genetics, a two allele model with irreversible mutations is an important limiting case in the theory of quasi-species \cite{eigen,quasispecies1,quasispecies2}.  In this theory, the phase transition corresponds to an ``error threshold'' at which a well adapted population's fitness distribution near a single fitness maximum becomes delocalized due to the large phase space available for deleterious mutations (see Appendix~\ref{AMutSelBal} for more details).
In this section, we will compare the well studied DK model results on a hexagonal lattice \cite{domany,lubeck,Hinrichsen,NEQPTBook} with our radial range expansion model results on a disordered Bennett lattice.

    When selection is strong, the green strain is able to survive even when $\mu_f > 0$.  However, for large enough $\mu_f$, we expect a ``mutational meltdown''  such that the advantageous green strain eventually goes extinct \cite{Hinrichsen, nelsonhallatschek}. 
It is known that in the biologically relevant region of reasonably small, nonzero deleterious mutation rates (e.g. $0<\mu_f< 0.5$),  the model exhibits a phase transition in the directed percolation universality class.   This transition is in the one-dimensional compact directed percolation (CDP) or ``voter model'' universality class when $\mu_f = \mu_b = 0$.  The phase diagram as a function of parameters analogous to our parameters $s$ and $\mu_f$  for $\mu_b=0$  has been studied extensively  (see \cite{Hinrichsen, NEQPTBook} and references therein) and  is illustrated in  Fig.~\ref{fig:PhaseDiagram}.  Note that the CDP transition is connected to an entire line of directed percolation phase transitions.

Near the phase transition line, various dynamical quantities  follow power laws in time, similar to the $1/t$ behavior of the density of surviving green cells with random initial conditions found in mean field theory (Eq.~(\ref{eq:MFf})) \cite{Hinrichsen, NEQPTBook}.  In addition, for a fixed mutation rate $\mu_{f}>0$, one can show that there are four critical exponents that characterize the scaling of the system.  Two exponents, $\nu_{\perp}$ and $\nu_{\parallel}$, describe the diverging correlation lengths measured perpendicular and parallel to the timelike direction.  If the directed percolation transition occurs at a critical selective advantage $s_c(\mu_f)$, and $\delta = s- s_c$ $(0 < \delta \ll 1)$ measures the distance from the critical point in the active phase, then a typical width and length of the red cell regions diverge as $\xi_{\perp} \sim \delta^{-\nu_{\perp}}$ and $\xi_{\parallel} \sim \delta^{-\nu_{\parallel}}$, respectively.  In the inactive phase, the green sector width and length diverge with the same exponents as we approach the phase transition.

Two additional exponents, $\beta$ and $\beta'$, describe order parameters relevant for two distinct  initial conditions: (1) the average density of active sites (green cells) at long times, $ \rho_{\infty} \equiv \langle f(t \rightarrow \infty) \rangle$  starting from $f(0) =1; $ and (2) the long-time sector survival probability $S_{\infty} \equiv S_{\mathrm{ss}}(t \rightarrow \infty)$ of a single green cell (more generally, any small green sector) seeded in an all red homeland.  Here $S_{\mathrm{ss}}(t)$ is the probability that a single sector has survived out to time $t$.  We  then have $\rho_{\infty} \sim \delta^{ \beta}$  and $S_{\infty} \sim \delta^{\beta'}$ for small distances $\delta>0$ into the active phase. In the biological context, the exponent $\beta$ describes the mutational meltdown  in a population of green cells during a range expansion, and $\beta'$ characterizes the survival probability of a rare advantageous mutant that spreads through the population while experiencing deleterious back mutations to a less fit  strain.    See Fig.~\ref{fig:DKPhase} for $\rho_{\infty}(\mu_f,s)$ for radial expansions.  Although we don't expect a sharp phase transition (see below), the general characteristics mimic the linear expansion diagrammed in Fig.~\ref{fig:PhaseDiagram}.

An important feature of the directed percolation (DP) dynamics described by Eqs.~(\ref{eq:LangevinVMfull}, \ref{eq:LangevinCorrVMfull}) is the so-called \textit{rapidity reversal} symmetry that implies $\beta = \beta'$.  Rapidity reversal is a special kind of time reversal that can be seen explicitly in the field theoretic formulation of the Langevin equation, which is known to be equivalent to a Reggeon field theory \cite{Hinrichsen,NEQPTBook}.  Rapidity reversal symmetry is only valid asymptotically  as $\delta \rightarrow 0^+$, where it implies that $\rho(t) \equiv \langle f(t) \rangle \propto  S_{\mathrm{ss}}(t)$ near the directed percolation phase transition for large values of $t$.  Indeed, both quantities scale for long times like $\rho(t) \sim S_{\mathrm{ss}}(t) \sim t^{-\alpha}$, where $\alpha = \beta/\nu_{\parallel} = \beta'/\nu_{\parallel}$. The asymptotic relation $\rho(t) \propto  S_{\mathrm{ss}}(t)$  is also valid for systems with time-independent, small system sizes since the size is invariant under time reversal (see Ref.~\cite{finitesize} for an example).      However, unlike a typical finite size effect, inflation is inherently asymmetric in time, and our simulations will show that inflation breaks the rapidity reversal symmetry, leading to new physics beyond conventional finite size scaling.

\begin{figure}
\includegraphics[height=2.3in]{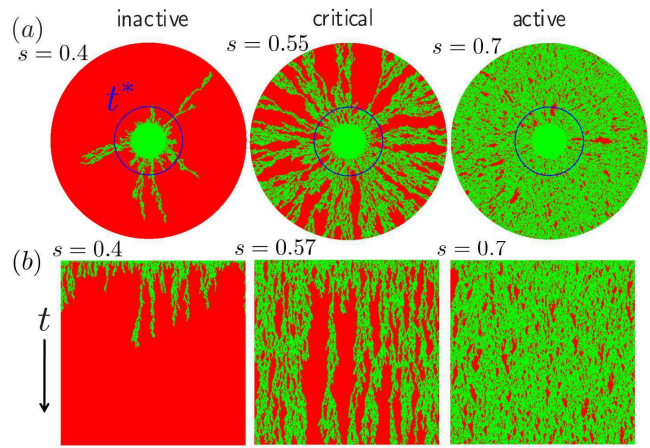}
\caption{\label{fig:phases} (Color online) Linear and radial range expansions with $\mu_b = 0$ and $\mu_f = 0.1$ for various $s$ near the directed percolation phase transition.  Green (light gray) organisms have a selective advantage $s$, but are subject to forward mutations to a less fit red (dark gray) strain.  In the radial expansions $(a)$  $R_0=50$ and the system evolved  for $\sim 250$ generations. The dark blue solid circles indicate the crossover time $t^* = 50$.   In $(b)$ a linear range expansion of $300$ cells evolves for $400$ generations.   }
\end{figure}

\begin{figure}
\includegraphics[height=2in]{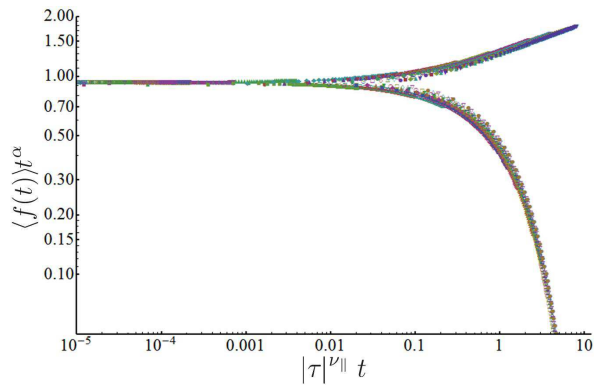}
\caption{\label{fig:dencollapse} (Color online) Data collapse for a radial range expansion with $R_0 = 2048$ for the average number of green cells over about 600 generations.  We test many values for $\tau = s-s_c = \pm 0.01,0.02,\ldots$ for $\mu_f = 0.1$.  The collapses are consistent with the directed percolation values $\alpha \approx 0.159464(6)$ and $\nu_{\parallel} \approx 1.733847(6)$. The critical selection parameter  was approximately $s_c \approx 0.55(1)$.}
\end{figure}

Consider the DP transition in the radial setting.  The plot in Fig.~\ref{fig:DKPhase} of the averaged green cell density after 1800 generations $(\rho(t=1800) \approx \rho_{\infty})$ for an all green homeland with $R_0=200$ shows that we have a crossover between an active and an inactive phase.     Despite the existence of these two phases, interpreting the critical point for radial expansions as a DP phase transition requires care!  Inflation  inhibits the boundaries of  green sectors that have formed during time $t^*$ from interacting diffusively at longer times $t \gg t^*$;  hence a large enough green sector could, with small probability, survive to the inflationary regime even in the inactive phase, after which it will persist for a very long time.  Of course, increasing the forward mutation rate $\mu_f$  still creates a crossover in the radial expansion dynamics, since any green sector will eventually become contaminated with red mutations.  

 The crossover in a radial range expansion is slightly shifted relative to the linear expansion transition line (dashed line in Fig.~\ref{fig:DKPhase}).
 The precise value of the transition is not universal  and varies with the lattice model details.  The direction of the shift in our case might be explained by the larger effective coordination number $z_r>2$ in the disordered Bennett lattice  relative to the hexagonal lattice: A larger $z$   should move the transition closer to the mean field line (dotted line in Fig.~\ref{fig:DKPhase}) as each daughter in a new generation can evolve from more potential parents.

  For early times $t \ll t^*$, inflation should only have a minor effect on  the coarsening dynamics at the transition.   As a check, Fig.~\ref{fig:dencollapse} shows a collapse of the data for the average fraction $\rho(t) = \langle f(t) \rangle$ of green cells for a variety of selective advantages  starting  with an all green initial population with $R_0 = 2048$. Our simulations run for about 600 generations for different values of $s$ at a fixed $\mu_f = 0.1$.   We successively collapsed the data using the remarkably precise estimates for the directed percolation exponents $\alpha \approx 0.159464(6)$ and $\nu_{\parallel} \approx 1.733847(6)$ quoted in Ref.~\cite{Hinrichsen}.  As expected from Fig.~\ref{fig:DKPhase}, the position of the apparent critical point for times $t \ll t^*$ is shifted slightly relative to the linear case ($s_c \approx 0.55$ versus $s_c \approx 0.57$ for $\mu_f = 0.1$).  Fig.~\ref{fig:dencollapse} confirms that the early time dynamics of our model is indeed accurately governed by directed percolation. 

At late times $t \gg t^*$,  the population fronts become locally flat due to the increasing population radius.  Thus, one might naively expect that the $t \gg t^*$ dynamics at the transition is also governed by critical DP dynamics.  However, inflation causally disconnects portions of the population after time $t^*$.  Thus, the $t \gg t^*$ dynamics consists of a set of noninteracting  DP processes.  The  fluctuations around time $t^*$ set the initial conditions for each process.  These initial conditions can be quite different (ranging from all green cells to a single green cell) and lead to different DP dynamics with different power laws \cite{Hinrichsen, NEQPTBook}.   Since we average over the behaviors of all the sectors in a radial range expansion,    the  $t \gg t^*$ regime does not correspond to the usual DP dynamics with a single initial condition.

One case in which the long-time behaviors of the linear and radial range expansions \textit{are} similar is in the limit $\mu_f \ll s$ for an all green initial condition, in which the expansions both achieve an active steady state.  The dynamics is deep in the ``active'' phase -- mutations generate small red sectors which die out with high probability.  If these sectors are sufficiently dilute, we can neglect their interactions.    This phase is illustrated on the right panel in Fig.~\ref{fig:phases}$(a)$.  Selective advantage bias tends to squeeze out red sectors, which we model by first replacing $w$ by $-w$ in  Eq.~(\ref{eq:ssFPE})  to study the probability distribution of \textit{red} sector widths. However, unlike the analysis of  green sector widths in Sec.~\ref{SSOthers},  the ``homeland'' radius $R_0$ must now be replaced by $R(\overline{t})=R_0+v \overline{t} > R_0$, where $\overline{t}$ is the time at which a red sector is generated by mutations.

To obtain the long-time behavior of the density $\rho_{\infty}\equiv \rho(t \rightarrow \infty)=\langle f (t \rightarrow \infty) \rangle$ of green sites,  we examine the survival rate of red sectors formed at very large times $\overline{t}\gg t^*$.  In this case, the fate of the red sector is only weakly influenced by inflation.  The red sector size probability distribution then obeys
\begin{equation}
\partial_{t} p_s(\phi, t) =  \overline{w} \, \frac{\partial p}{\partial \phi} + \overline{\Delta} \, \frac{\partial^2 p}{\partial \phi^2}, \label{eq:redsecdiff}
\end{equation}
with the \textit{constant} coefficients, for a fixed time $\overline{t}  \gg t^*$,
\begin{equation}
\overline{w}\approx \frac{ \gamma s  \tilde{a}_r}{\tau_g R(\overline{t})} \mbox{\quad and \quad} \overline{\Delta} \approx \frac{\tilde{a}_r^2 }{\tau_g(R(\overline{t}))^2}. \label{eq:redsectorparams}
\end{equation}
 As usual, we impose an absorbing boundary condition $p(\phi=0,t)=0$ on Eq.~(\ref{eq:redsecdiff}).  If we change variables to make Eq.~(\ref{eq:redsecdiff}) a conventional diffusion equation, then the boundary condition now \textit{advances} on the domain where diffusion and selection bias take place.  In such a case, the absorbing state (extinction of a red sector) is always reached \cite{redner}.  Finally, note that we are no longer strictly in the small $s$ regime and hence have no simple expression for $\gamma$ in Eq.~(\ref{eq:redsectorparams}). However, we expect it to be close to unity (see Appx.~\ref{App:SingleSectorInfl}). Since we are only interested in an  approximation of the scaling of $\langle f (t \rightarrow \infty) \rangle$ with $\mu_f$ and $s$,  we set $\tilde{a}_r=\gamma=1$ in the following for simplicity.  

\begin{figure}
\includegraphics[height=2in]{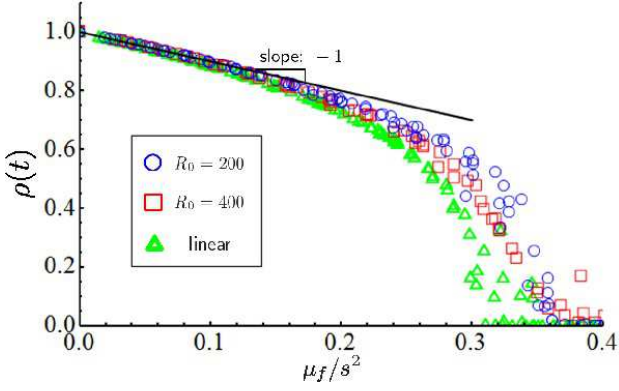}
\caption{\label{fig:APcollapse} (Color online) We plot the average green cell density $\rho(t) \equiv \langle f(t) \rangle$ at  long times $t$ versus a control parameter $\mu_f/s^2$ for an initially all green population  for radial and linear expansions.  For radial expansions, $f(t)$ is averaged over $10^{3}$ runs at $t = 2 \times 10^3$ generations. In the linear case, we averaged $f(t)$ over $2 \times 10^3$ runs, at time $t = 2 \times 10^4$, for a system size of $10^3$. Away from the transition region ($\mu_f/s^2 \lesssim 0.2$), the time $t$ is long enough so that the density $\rho(t)$ for both expansions is approximately equal to the steady-state density $\rho_{\infty} \equiv \langle f (t \rightarrow \infty) \rangle$, with an error smaller than the point size.   The black line shows the predicted scaling of the steady-state density for small $\mu_f/s^2 $.  We varied  $s$ between $0$ and $1$ and $\mu_f$ between $0$ and $0.5$.}
\end{figure}

The average excursion area (red sector size) $\langle A_{\mathrm{red}} \rangle$ of the random walk described by Eq.~(\ref{eq:redsecdiff}) before it reaches the absorbing state,  to first order in the initial sector size of one cell, $\phi_0 \approx  \tilde{a}_r/R(\overline{t})$,  is given by a known result from diffusion theory (see e.g., Ref.~\cite{kearney}):
\begin{equation}
\langle A_{\mathrm{red}} \rangle = \frac{\overline{\Delta} R(\overline{t}) \phi_0}{(\overline{w})^2} \approx\frac{1  }{ s^2 } . \label{eq:redexarea}
\end{equation}
Next, consider a population of green cells (in the active state) evolving from time $t=\overline{t} $ to time $\overline{t}+T$ (with $\overline{t} \gg T,t^*$). The number of red sectors $N_{\mathrm{red}}$ that are seeded in the population during this time will be approximately $N_{\mathrm{red}} \approx 2 \pi \mu_f vTR(\overline{t})/ \tilde{a}_r^2$.  The total area covered by the population from time $\overline{t}$ to time $\overline{t}+T$ is  $A_{\mathrm{tot}} \approx 2 \pi R(\overline{t}) vT$.  Upon dividing the average area covered by the green cells by $A_{\mathrm{tot}}$, we find the steady-state density  of the green cells:
\begin{eqnarray}
\rho_{\infty} & \equiv & \langle f (t \rightarrow \infty) \rangle  \nonumber \\
& = & \frac{A_{\mathrm{tot}}- N_{\mathrm{red}}\langle A_{\mathrm{red}} \rangle }{A_{\mathrm{tot}}} \approx 1- \frac{\mu_f  }{  s^2 } \quad (\mu_f \ll s). \label{eq:APscaling}
\end{eqnarray}
Note the crucial difference between Eq.~(\ref{eq:APscaling}) and the steady state value $1-\mu_f/s$ in the mean field (well mixed) solution in Eq.~(\ref{eq:MFf}):   For fixed values of $\mu_f$, we require  larger values of $s \approx \sqrt{\mu_f} > \mu_f$ to get a comparable reduction in $\rho$ for the circular range expansion.

 Eq.~(\ref{eq:APscaling}) is derived in the absence of inflation, so the  same expression for $\rho_{\infty}$ holds for linear range expansions, similar to the result obtained for a related model in Ref.~\cite{nelsonhallatschek}.  This is consistent with the similarity between the active phase pictures for radial and linear range expansions in Fig.~\ref{fig:phases}$(a)$ and $(b)$, respectively.   We now check Eq.~(\ref{eq:APscaling}) against our data in the phase diagram of Fig.~\ref{fig:DKPhase} for both linear and radial expansions.   Fig.~\ref{fig:APcollapse} shows collapse of the data for $\mu_f/s^2 \ll 1$, i.e., deep in the active phase.  As argued in Ref.  \cite{nelsonhallatschek},  $\rho_{\infty}$ is driven below $1 - \mu_f/s^2$ by collisions between red sectors, which merge to shield green sectors from invading the red regions, increasing the total fraction of red cells.   Note that the slope of the line at small $\mu_f/s^2$ in Fig.~\ref{fig:APcollapse} is consistent with our  approximation $\tilde{a}_r = \gamma = 1$.

As we keep increasing $\mu_f/s^2$, the steady-state density $\rho_{\infty}$   eventually goes to zero.  In the linear case, there is a sharp transition at around $\mu_f/s^2 \approx 0.3$, consistent with the results of Ref.~\cite{nelsonhallatschek}.  This extinction of the green organisms is triggered, of course, by crossing the directed percolation phase transition line \cite{Hinrichsen}.  A sharp transition is less evident in Fig.~\ref{fig:APcollapse} for \textit{radial} expansions:  the density $\rho_{\infty}$ seems to decay to zero more slowly as we approach the crossover.   However, our estimate of $\rho_{\infty}$ in Fig.~\ref{fig:APcollapse} may not be reliable close to the crossover as the system takes a long time to reach a steady state in this region.    We shall now study the dynamics of the radial range expansion near this crossover.

\begin{figure}
\includegraphics[height=3.6in]{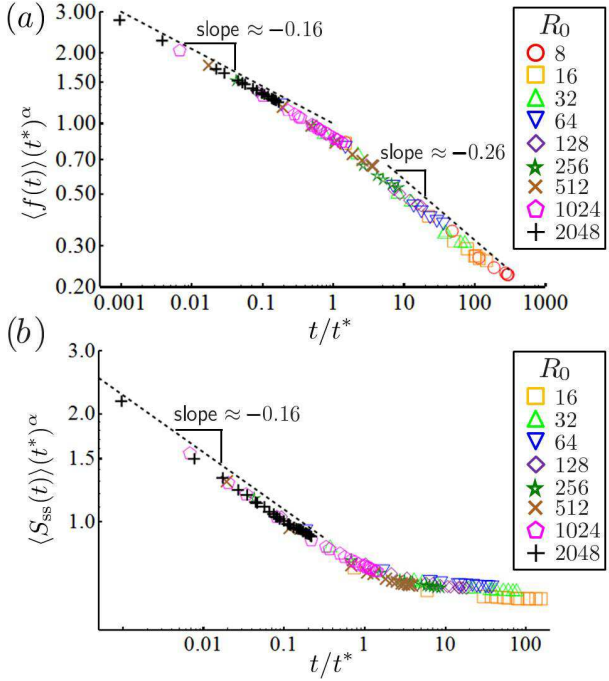}
\caption{\label{ffinitesize} (Color online) Data collapses for radial range expansion simulations with many values of $t^* = R_0/v$ (with $\mu_f = 0.1$ and $s = 0.547$).  In (a) we plot the scaled green cell density $\rho(t) = \langle f(t) \rangle$  with an all green initial condition for $R_0 = 8,16,32,\ldots,2048$.  The regimes $t \ll t^*$ and $t \gg t^*$ can be described by power laws.  The  $t^{-\alpha}$ decay for $t < t^*$  with $\alpha \approx 0.16$ is expected from directed percolation theory \cite{Hinrichsen}, but the exponent $\alpha \approx 0.26$ that arises for long times appears to be new.   In (b) we plot the scaled sector survival probability $S_{\mathrm{ss}}(t) $ of a single green cell inoculated in an all red homeland $R_0 = 16,32,\ldots,2048$.  The dashed line indicates the expected directed percolation result: a decay with slope  $-\alpha \approx -0.16$.  Note, however, that the survival probability levels off to a finite value for $t \gg t^*$ due to inflation.   }
\end{figure}

To explore directed percolation dynamics when inflation has set in, i.e. for $t > t^* = R_0/v$,  we use a dynamic scaling hypothesis for the average green fraction $\rho(t)=\langle f(t) \rangle$ and for $ S_{\mathrm{ss}}(t) $, where $S_{\mathrm{ss}}(t)$ is the survival probability of a single green cell inoculated at the edge of a red homeland.   We let $\delta = s-s_c(\mu_f)$ be the distance from the linear inoculation $(R_0 \rightarrow \infty)$ critical point.  We approximate the critical point $s_c(\mu_f)$ by looking at the short time $(t \ll t^*)$ dynamics of a radial range expansion with a large homeland radius $R_0$.  For example, we found that $s_c(\mu_f=0.1) \approx 0.55$ (see Fig.~\ref{fig:dencollapse}).  Note that we cannot use the hexagonal lattice for linear expansions to find the critical point because we expect the point to shift when we use the disordered Bennett lattice. 

Inspired by finite size scaling ideas near critical points, we introduce a scaling transformation $\delta \rightarrow b \delta$ and expect the scaling relations
\begin{align}
\rho(t)= b^{-\beta} \tilde{\rho} \left[ b^{-\nu_{\parallel}} t, b \delta,b^{-\nu_{\parallel}}t^*\right], \\
 S_{\mathrm{ss}}(t)  = b^{-\beta'} \tilde{S}_{\mathrm{ss}} \left[ b^{-\nu_{\parallel}} t, b \delta,b^{-\nu_{\parallel}}t^*\right],
\end{align}
where we have suppressed metric factors \cite{Hinrichsen}, and $\tilde{\rho}$ and $\tilde{S}_{\mathrm{ss}}$ are scaling functions.  We now set $b^{-\nu_{\parallel}}t^* = 1$ and find that our data for various $t^*$  (corresponding, say, to different homeland radii $R_0 = v t^*$) at  the critical point ($\delta = 0$) should collapse upon plotting $\rho(t) ( t^*)^{\alpha }$ and $ S_{\mathrm{ss}}(t)  (t^*)^{\alpha }$ versus $t/ t^*$.  These scaling forms are tested in Fig.~\ref{ffinitesize}.  The data collapse reasonably well,  and  we indeed get a crossover at $t/t^* \approx 1$ in both cases.  The inflation crossover time $t^*$ enters just as a finite size effect would in conventional critical phenomena.

Both of the collapses are consistent with a single value $\alpha$ given by  directed percolation theory \cite{Hinrichsen}.  This suggests that we retain the rapidity reversal symmetry of  directed percolation in a linear geometry at early times.    However, the rapidity reversal prediction \cite{Hinrichsen} $\rho(t) \propto  S_{\mathrm{ss}}(t)$ is violated in the inflationary regime $t > t^*$.   Inflation  prevents green sectors from dying out for a  single seed initial condition (Fig.~\ref{ffinitesize}b), and the survival probability $S_{\mathrm{ss}}(t)$ approaches a nonzero constant at long times.  Conversely, inflation  evidently enhances the loss of the total fraction of green cells for the  green homeland initial condition, and the density $\rho(t)$ decays to zero as $t \rightarrow \infty$ (Fig.~\ref{ffinitesize}a).   In a conventional finite size effect, $S_{\mathrm{ss}}(t)$ and $\rho(t)$ are consistent with rapidity reversal symmetry, and both decay exponentially to zero at long times \cite{finitesize}.

The levelling off of the survival probability $S_{\mathrm{ss}}(t)$ in Fig.~\ref{ffinitesize}$(b)$ for $t \gg t^*$ is consistent with the analysis of single sectors without mutations in Sec.~\ref{SSOthers}:  The survival probability of a single sector approaches a constant at long times (see Eq.~\ref{eq:limitingsurvpwsel}) even at $s = 0$.   Unlike a sector in a linear expansion, the diffusive motion of the sector boundaries in a radial range expansion is suppressed in the inflationary regime.  The boundaries are unlikely to collide in this regime, and a sector that makes it to the inflationary regime can survive even as $t \rightarrow \infty$.        In a linear expansion, $S_{\mathrm{ss}}(t)$ decays to zero at the directed percolation transition as $t^{-\alpha}$ for large times $t$.      

The more rapid decay of the average fraction of green cells $\rho (t)$ in Fig.~\ref{ffinitesize}$(a)$ for $t > t^*$ is more subtle: $\rho (t)$ seems to decay approximately as $t^{-0.26}$.  We can understand this trend qualitatively by appealing to the mean field solution for $\rho (t)$ (see Eq.~(\ref{eq:MFf})) which decays at the much faster rate $t^{-1}$ at criticality.  Since inflation suppresses both diffusion and noise in our Langevin equation for $f(\mathbf{x},t)$ (Eq.~(\ref{eq:RadLangevin})), it is plausible that $\rho (t)=\langle f(t) \rangle$ decays  faster than for the linear expansion where diffusion and noise are more important.  However, it is not clear that the larger decay exponent in the inflationary regime is universal, and we do not have an explanation for its particular value of $-0.26$.  

Another important dynamical quantity is the pair connectedness correlation function \cite{NEQPTBook}, i.e., the probability $\Upsilon(x,t ;x_0,t_0 )$ of finding a directed,  causal chain of entirely green cells between a green cell at point $(x_0,t_0)$ and  any point $(x,t)$ for $t > t_0$.   For radial Domany-Kinzel expansions, $\Upsilon(x,t ; x_0,t_0 )$ is the probability that a green cell at position $x_0 = R(t_0) \phi_0$ along the population circumference at time $t_0$ has a green cell descendant at some later time $t$ at position $x=R(t)\phi$.  Thus, $\Upsilon(x,t ;x_0,t_0 )$ characterizes the spatial correlations of genetic lineages in the green cells.      The pair connectedness function does not keep track of the red cell lineages, including those red cells that evolved from green cell ancestors. A knowledge of $\Upsilon(x,t ;x_0,t_0 )$ might be useful for genetic inference, where spatial distributions of genetic variants at time $t$ are used to infer ancestral genetic distributions (see Ref.~\cite{KorolevRMP} for more information). 

We can probe $\Upsilon(x,t ;x_0,t_0 )$ by employing a single green cell initial condition at the origin and monitoring the resultant green sector formed by descendants of the single green ancestor.  We compute the total number of green (i.e. active) cells $N_a(t)$ at time $t$ and the mean  square spread of green clusters that survive to time $t$:  $X_a^2(t) \equiv \langle |\Delta x|^2 \rangle$,  where 
\begin{equation}
 \langle |\Delta x|^2 \rangle = \frac{\int (\Delta x)^2 \Upsilon \left(\Delta x,t; 0,0 \right) \, \mathrm{d} (\Delta x)}{\int \Upsilon(\Delta x,t; 0,0) \,\mathrm{d} (\Delta x)},
\end{equation}
and we average over all green cells located at displacements  $\Delta x = R(t) \Delta \phi$ away from the  initial green cell. The quantity   $N_a(t)$ is related to $\Upsilon(x,t ; 0, 0)$ via  \cite{NEQPTBook}
\begin{align}
 N_a(t) & = \mbox{const.} \times \int \mathrm{d} x \, \Upsilon(x,t;0,0)
\end{align}
and obeys a scaling relation (see Ref.~\cite{Hinrichsen,NEQPTBook})
 \begin{equation}
N_a(t)  = b^{\nu_{\perp}-\beta-\beta'} \tilde{N}_a(b^{-\nu_{\parallel}}t, b  \delta,b^{-\nu_{\parallel}}t^*).
\end{equation}
 Similarly, we have
\begin{align}
 X_a^2(t) & = \frac{\mbox{const.}}{N_a(t)} \int \mathrm{d} x\,|x|^2 \, \Upsilon(x,t;0,0) 
\end{align}
and the scaling relation
\begin{equation}
X_a^2(t) = b^{2 \nu_{\perp}}  \tilde{R}^2_a(b^{-\nu_{\parallel}}t, b \delta,b^{-\nu_{\parallel}}t^*),
\end{equation}
where we have again suppressed metric factors \cite{NEQPTBook}.

\begin{figure}
\includegraphics[height=3.6in]{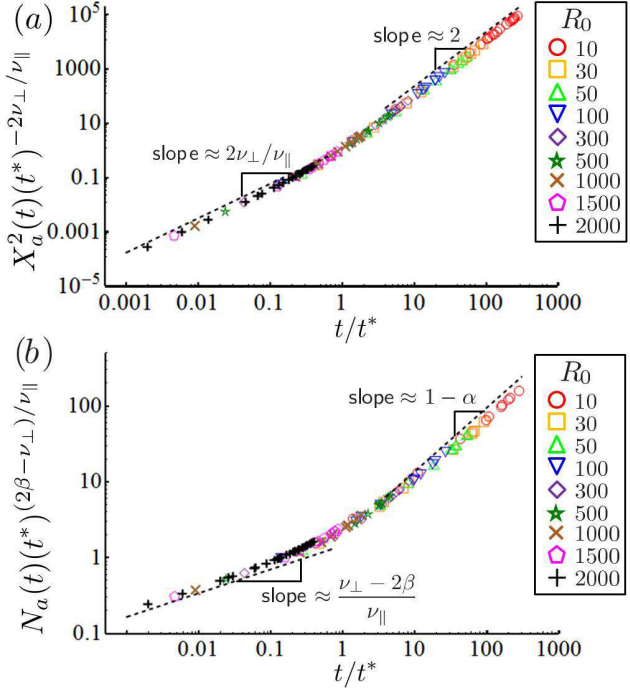}
\caption{\label{fig:pairconnect} (Color online) Data collapse for radial expansion simulations with a single green cell  in an otherwise red initial homeland perimeter (with $\mu_f = 0.1$, $\mu_b = 0$, and $s = 0.547$) on log-log plots.    $(a)$ The rescaled mean squared cluster spread $ X^2_a(t) $. The dashed line for $t \ll t^*$ indicates the directed percolation power law scaling with exponent: $2 \nu_{\perp}/\nu_{\parallel} \approx 1.3$. The dashed line in the inflation-dominated $t \gg t^*$ regime agrees with a $R_a^2(t) \sim t^2$ behavior discussed in the text.   $(b)$ Plot of the rescaled average number of green cells $ N_a(t) $.  The  expected directed percolation scaling for $t \ll t^*$ is shown with a dashed line, with $(\nu_{\perp}-2 \beta)/\nu_{\parallel} \approx 0.31$.  For $t \gg t^*$, we find  $N_a(t) \sim t^{1-\alpha} \approx t^{0.84}$, consistent with arguments incorporating inflation given in the text. The simulation consisted of $9.6 \times 10^4$ independent runs for each homeland with radii $R_0 = 10,30,50,100,\ldots,2000$, evolved to a final population of size $R(t) = 2800$.}
\end{figure}

Upon setting $b^{-\nu_{\parallel}}t^* = 1 $, we now find data collapse when we plot $X_a^2(t) (t^*)^{-2 \nu_{\perp}/\nu_{\parallel}}$ and $N_a(t) (t^*)^{(\beta+\beta'-\nu_{\perp})/\nu_{\parallel}}$ versus $t/t^*$ at criticality $(\delta = 0)$ (see Fig.~\ref{fig:pairconnect}). In our simulations, $X_a^2(t)$ can be computed by finding the average squared angular spread $\langle \phi^2 \rangle$ and multiplying by $R^2(t)=(R_0+vt)^2$. We find excellent data collapses for both quantities.  Notice that for times $t \ll t^*$, we get the expected power-law behavior given by the directed percolation universality class \cite{NEQPTBook}.

For $t \gg t^*$, we have argued that a green sector evolved from a single green cell will expand deterministically.  The angular width of the sector will increase somewhat due to the selective advantage, but inflation will be the  dominant effect at long times. Hence, the main contribution to $X_a^2$ for $t \gg t^*$ arises from large green sectors evolving with fixed angular widths.  We then expect $X_a^2(t) \sim t^2$   at long times, consistent with the long-time behavior in Fig.~\ref{fig:pairconnect}$(a)$.  The main contribution to $N_a(t)$ also arises from this deterministic expansion epoch.  However, we expect the density of green cells \textit{inside} each sector to decay according to  $t^{-\alpha}$, since the interior of a green sector should be describable as a critical directed percolation process.  Thus, we expect that $N_a(t) \sim t^{1-\alpha}$ for $t \gg t^*$.  This expectation also matches the simulation results (see Fig.~\ref{fig:pairconnect}$(b)$).  By contrast, directed percolation in a finite geometry is unable to sustain large cluster growth.  Thus, at long times, $N_a(t)$ and $X_a^2(t)$ have different scaling functions that rapidly decay to zero \cite{finitesize}.

The scaling result for the mean square spread $X_a^2$ of surviving green clusters  suggests a simple prediction for directed percolation with arbitrary power law inflation $R(t) = R_0[1+(t/t^*)^{\Theta}]$.  Namely, for $\Theta > \nu_{\perp}/\nu_{\parallel} \approx 0.632$, the inflation should be able to overtake the DP critical cluster growth, so that $X_a^2 \sim t^{2 \Theta}$ and $N_a \sim t^{\Theta-\alpha}$ at long times $t \gg t^*$.  $X_a^2$ and $N_a$ should still have their regular DP power laws for $t \ll t^*$. However, the shape and position of the crossover region at $t \sim t^*$ will likely change with $\Theta$.   Conversely, when $\Theta < \nu_{\perp}/\nu_{\parallel}$, the cluster growth is always faster than the inflation, and we expect the inflation to be \textit{irrelevant} so that $X_a^2 \sim t^{2 \nu_{\perp}/\nu_{\parallel}}$ and $N_a \sim t^{(\nu_{\perp}-2 \beta)/\nu_{\parallel}}$  at long times for all $\Theta < \nu_{\perp}/\nu_{\parallel}$.    Of course, a rigorous check of this simple prediction is necessary. It would also be interesting to study the borderline case $\Theta = \nu_{\perp}/\nu_{\parallel}$.

\section{\label{SSReverseInflation}Radial Range Expansions with Deflation}

\begin{figure}
 \includegraphics[height=2.1in]{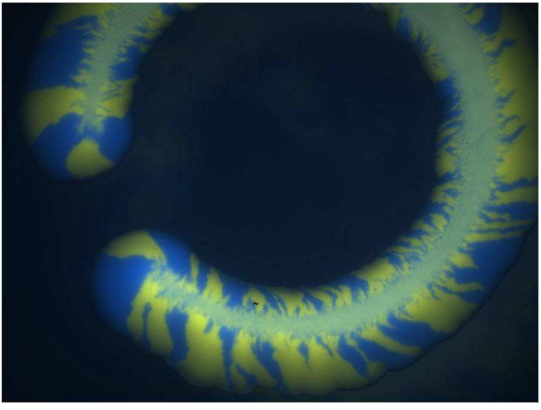}
\caption{\label{fig:bacexp}  (Color online) An inoculation of two bacterial (\textit{Escherichia coli}) strains on a Petri dish, labelled with blue (dark gray in printed micrograph) and yellow (light gray) constitutively expressed fluorescent proteins.  The strains are otherwise identical and, consequently,  represent the $s = \mu_f = \mu_b = 0$ case in our analysis.  The initial inoculation has the shape of a partial ring to allow for nutrients to penetrate the ring center and allow for the bacteria to grow inward, creating a ``deflating'' radial range expansion.  A regular ``inflating'' radial range expansion moves away from the ring.  This single experiment illustrates both kinds of radial expansion discussed in this paper! }
\end{figure}

Another type of radial expansion starts with a population around the circumference of a circle of radius $R_0$.  The population then grows \textit{inward} toward the center of the circle, colonizing the interior.  An example of such a ``deflating'' range expansion is the settling of an island by a new species  that arrives simultaneously around the shoreline and then grows inland.    The population front at time $t$ in this case is a circle of radius $R(t)=R_0-vt$.  Note that $t^* = R_0/v$ is now a sharp temporal \textit{cutoff} (instead of a crossover), at which the population front size vanishes and the invading species completely takes over the circular island.  

 A deflating range expansion can be realized in experiment by inoculating a ring of organisms on a Petri dish, as shown in Fig.~\ref{fig:bacexp}.  As long as there are enough nutrients on the interior of the ring, the organisms will grow inward toward the ring center.   Fig.~\ref{fig:bacexp} shows the genetic demixing occurring between two neutral bacterial strains labelled with constitutively expressed blue and yellow fluorescent proteins (dark and light gray regions, respectively, in printed Fig.~\ref{fig:bacexp}).   The bacterial sectors   exhibit a counterclockwise ``twisting''  that is analyzed in detail in Ref.~\cite{KorolevBac}.  This uniform tangential motion of the bacterial sector boundaries does not affect the genetic demixing or quantities such as the heterozygosity that only depend on the separation between two individuals along a population front.  We will not model the twisting in simulations, but note that it can be easily implemented by shifting all of the cells at a population front counterclockwise at each time step.

\begin{figure}
\includegraphics[height=1.9in]{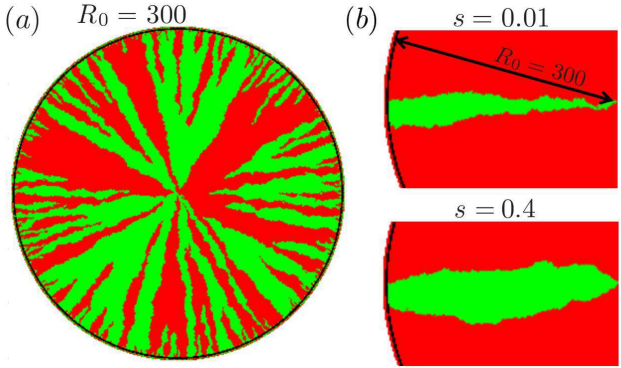}
\caption{\label{fig:deflationex} (Color online) $(a)$ A radial range expansion with a deflating  population front, for $s = \mu_f = \mu_b = 0$, starting from an initial population around a circular rim of radius $R_0 = 300$.  We use the same Bennett model lattice described in Sec.~\ref{SSimulations}, while running the update scheme in reverse.  $(b)$ Deflationary range expansions for a single  green (light gray) sector with an initial size $\phi_0 = 0.1$ radians for a large ($s=0.4$) and very small ($s=0.01$) selective advantage.  In both cases, deflation forces the  green (light gray) sector boundaries to collide at the center (when $t = t^* = R_0/v$).    }
\end{figure}

The simulated evolution of two neutral strains in a deflating range expansion is shown in Fig.~\ref{fig:deflationex}$(a)$.  A well mixed initial condition is used with equal proportions of green (light gray) and red (dark gray) cells.  The initial population is on a circular boundary with radius $R_0 = 300$.  The deflationary dynamics, similar to the inflationary dynamics considered in Sec.~\ref{SSNeutralEvo}, is related to a linear range expansion via a conformal time coordinate $t_c(t) = t/(1-t/t^*)$. The heterozygosity is again given by   $H(\phi,t)= H_0\, \mbox{erf} \, |\xi|$, with $\xi = R_0 \phi/\sqrt{8 D_r t_c(t)}$. However, unlike the inflationary case, $H(\phi,t)$ now decays to zero and becomes flat as $t \rightarrow t^*$!  Thus, the deflating range expansion  evolves to a population with just one genetic sector as $t \rightarrow t^*$, as shown in  Fig.~\ref{fig:deflationex}$(a)$.

The dynamics of a single green sector in a deflating radial range expansion (shown in Fig.~\ref{fig:deflationex}$(b)$)    also differs from the inflating case.  Using the diffusion equation techniques of Sec.~\ref{SSOthers} and App.~\ref{App:SingleSectorInfl}, we can calculate the deflationary survival probability $S_d(t)$ of a green sector with a small selective advantage $w$ and a small initial size $\phi_0 \ll 2 \pi$.  The result as $t \rightarrow t^*$ is\begin{align}
S_d(t \rightarrow t^*) &  \approx1- \int^{\infty}_{0}    \frac{e^{-x} \,}{  \sqrt{  \pi x}}    \,e^{- \frac{2w x\phi_0}{ 4 \Delta x+\phi_0^2}- \frac{w^2 x\phi_0^2}{ (4 \Delta x+\phi_0^2)^2}} \,\mathrm{d} x, \mbox{ or }   \label{eq:survpreverse1}\\
S_d(t \rightarrow t^*)& \approx \left[\frac{\phi_0}{ \Delta }-\frac{\sqrt{\pi } \phi_0^2}{2 \Delta ^{3/2}}\right] \frac{ w}{2 }+\left[\frac{\sqrt{\pi } \phi_0}{4 \Delta ^{3/2}}-\frac{\phi_0^2}{ \Delta ^2}\right] \frac{w^2}{4} \nonumber \\
& \quad   {}+\mathcal{O}(\phi_0^2w^3)+\mathcal{O}(\phi_0^3w) \quad (w,\phi_0 \ll 1). \label{eq:survpreverse2} \end{align}
Note that the convergence of sector boundaries insures that  $S_d(t \rightarrow t^*)$  vanishes in the absence of selection $(w=0)$.  This behavior differs markedly from the inflationary case in which $S_r(t \rightarrow \infty)$ has a limiting value given by  Eq.~(\ref{eq:limitingsurvp}) with $\tau = 1$.   However, when $t \rightarrow t^*$ in the deflationary case, the population size at the frontier goes to zero, and the finite size of the cells becomes important.  Hence, we do not expect the continuum diffusion model to hold exactly as $t \rightarrow t^*$; contrary to the continuum result (Eq.~(\ref{eq:survpreverse1})), a small residual survival probability is possible in this limit, especially for larger initial sector sizes $\phi_0$.  Thus, Eq.~(\ref{eq:survpreverse1}) and Eq.~(\ref{eq:survpreverse2})  are good approximations to the limiting survival probability only when the fate of the green sector is determined sufficiently far from the sector convergence time $t = t^*$.

\begin{figure}
\includegraphics[height=2.1in]{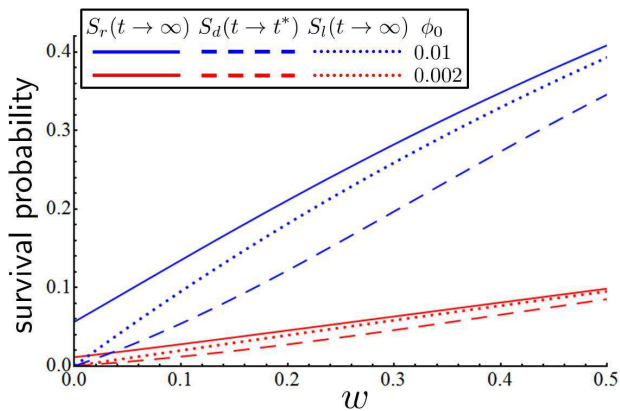}
\caption{\label{fig:survpcompare} (Color online) A comparison of the long-time survival probabilities  of green sectors for small selective advantages $w \propto s$ in range expansions with  radial inflation ($S_r(t \rightarrow \infty)$, solid lines), radial deflation ($S_d(t \rightarrow t^*)$, dashed lines), and for linear range expansions ($S_l(t \rightarrow \infty)$, dotted lines).  The probabilities are calculated using Eq.~(\ref{eq:radsurvint}), Eq.~(\ref{eq:survpreverse1}), and a similar  result  for a linear range expansion.  In the forward and reverse radial expansions, the initial population radius is $R_0 = 100 \, a_r$, where the effective lattice constant  $a_r=1$ for convenience.  The linear expansion assumes a total population size $L = 2 \pi R_0 = 200 \pi a_r$.   We used two different initial green sector sizes, $\phi_0 = 0.01$ (upper set of curves) and $\phi_0= 0.002$ (lower set of curves).  }
\end{figure}

 Comparing  Eq.~(\ref{eq:survpreverse2})  and Eq.~(\ref{eq:limitingsurvpwsel})  reveals that deflation suppresses the effects of selection relative to  inflation, an effect which arises because deflation forces the green sector boundaries closer together, as illustrated in Fig.~\ref{fig:deflationex}$(b)$.  The survival probabilities for inflationary and deflationary radial range expansions can also be compared to the limiting survival probability $S_l(t \rightarrow \infty)$ of a green sector in a linear range expansion with selective advantage $w$ and initial size $R_0 \phi_0$ in a population of linear size $L = 2 \pi R_0$.   The results for two different sector sizes $\phi_0$ are shown in Fig.~\ref{fig:survpcompare}.  As expected, the survival probability of a green sector in a linear range expansion  is bracketed by the probabilities for inflating and deflating radial expansions.

\section{\label{SConclusions}Conclusions}

This paper has explored differences between the population genetics of radial and linear range expansions in the presence of genetic drift, mutation, and selection.  We find that the inflation embodied in a radial expansion decreases the effect of genetic drift and biases the stochastic dynamics in a deterministic direction.  The proliferation of extra sites at the outer edge of the radial expansion greatly reduces sector interactions  after time $t^*=R_0/v$.  For times $t \gg t^*$, the population evolves in isolated angular segments, with boundaries perturbed by genetic drift.    Near the directed percolation (DP) phase transition, we found that $t^*$ acts as a finite crossover time, after which the radial expansion no longer experiences the critical dynamics.  Finite size scaling near the DP transition due to inflation is described by a different scaling function than that governing conventional finite size effects.

We presented and tested analytical approximations to the dynamics of single sectors in various limits, including selective advantages  weak compared to the genetic drift and inflation.  It would be interesting to develop a more precise asymptotic analysis of the radial survival probability $S_r(\phi_0,\tau)$ of a sector with an arbitrary initial angular width $\phi_0$ and $\tau = t/(t+t^*)$. We also calculated the survival probability in a deflationary radial range expansion in which the sector boundaries are pushed together by a collapsing population front.  We found that inflation enhances and deflation diminishes the survival probability relative to a linear range expansion with the same initial population size and selective advantage.

It would also be interesting to study the inflationary effects in more detail via a field theoretic derivation of the scaling hypothesis \cite{NEQPTBook, Hinrichsen}.  In the linear expansion, the relevant field theory is Reggeon field theory.  In the radial case, we would have to introduce the inflationary metric $d s^2 = v^2 \, d t^2+R(t)^2 \, d \phi^2$ (with front propagation speed $v=dR/dt$) for measuring distances between points in the  range expansion rather than the usual Euclidean one $d s^2 = v^2\, dt^2+ d x^2$ to take into account the expanding system  size.  So, we expect that the field theoretic description of radial expansions would involve studying the Reggeon field theory in this ``curved''  inflationary space.  

It would also be interesting to extend our results by allowing for undulations in the radial population front.  Such undulations produce small changes in the power laws predicted by simple diffusive models of sector growth and wandering \cite{nelsonhallatschek}.  One way to allow undulations in simulations is to generalize the Eden model proposed in Ref.~\cite{frey}  to an amorphous lattice and evolve initially circular homelands of individuals which  now divide freely at the population periphery.  Without the constraint that daughter cells always spawn as close as possible to the center of the expansion, the population front should roughen away from a nearly circular shape. In the future, we hope to study an extension of our model to three dimensions, thus allowing analysis of a greater variety of range expansions, such as tumor growth in cancer.  Spherical cell colonies, for example, can be modeled using the three-dimensional Bennett model \cite{Bennett}.  We expect that the effects of inflation are more pronounced in three dimensions since sector surface areas now increase quadratically in time due to inflation, while sector coarsening due to genetic drift is much slower at a two dimensional frontier \cite{KorolevRMP}.

\begin{acknowledgments}
We are pleased to acknowledge the support of the National Science Foundation, through Grant DMR 1005289, and through the Materials Research Science and Engineering Center, via Grant DMR 0820484.  M. O. L. acknowledges the support of the National Science Foundation Graduate Research Fellowship.  Computational support was provided by the National Nanotechnology Infrastructure Network Computation project and the Harvard SEAS Academic Computing group.
\end{acknowledgments}

\appendix

\section{\label{AMutSelBal}Quasi-species and Mutation-Selection Balance}

\begin{figure}
\includegraphics[height=1.8in]{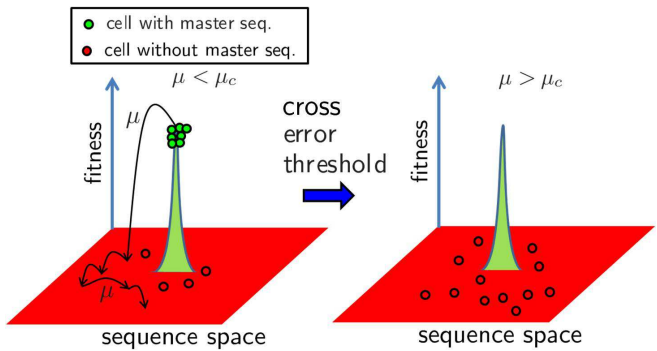}
\caption{\label{fig:quasispecies} (Color online) A schematic of the quasi-species model  in which each cell in a population has a sequence distributed in a fitness landscape with a single sharp peak  in  green (light gray) surrounded by a flat landscape in red (dark gray).  As cells divide, each individual reproduces with a mutational error  rate $\mu$ that moves it around in sequence space (black, thin arrows).  Below a critical ``error threshold'' $(\mu < \mu_c)$, a finite fraction of the population has the master sequence (i.e., is a green (light gray) cell). Cells deviating from this ideal sequence are red (dark gray). When $\mu > \mu_c$, the population fitness distribution  becomes delocalized, and the fitness peak is depopulated.  }
\end{figure}

An important biological motivation for studying irreversible, deleterious mutations is the molecular quasispecies theory developed by Eigen \cite{eigen}.  In this theory (reviewed in \cite{quasispecies1,quasispecies2}), each individual in an asexually reproducing population has a molecular sequence  (i.e., a DNA or a protein sequence).  During each reproduction step, the sequence gets replicated and acquires errors at some rate $\mu$.  The errors change the sequence's fitness. If the error rate is small, the population should become well adapted and its sequences will be clustered around a peak in a fitness landscape.  The individuals at the peak have a unique ``master sequence'' corresponding to this highest possible fitness in the population, as illustrated in Fig.~\ref{fig:quasispecies}.   Each mutation away from the master sequence is necessarily deleterious.  For a variety of fitness peaks with this general structure, there is a critical mutation rate $\mu_c$ above which the population fitness distribution becomes delocalized, and the fraction of individuals that preserve the master sequence goes to zero  at long times.  The critical mutation rate $\mu_c$ is called an ``error threshold'' \cite{quasispecies1}.

The irreversible mutation case $(\mu_b=0)$ we consider in Sec.~\ref{SSScaling} corresponds to a particularly simple fitness landscape  in which the fitness peak is infinitely sharp and each mutation away from the master sequence will have an identical deleterious effect with a selection coefficient $s$ (see  Fig.~\ref{fig:quasispecies} for an illustration and Ref.~\cite{quasispecies1} for more details) .  The green cells are the cells with the master sequence.  Cells without the master sequence are red and have the same fitness because  the landscape is flat away from the peak.  If the master sequence is very long, then the probability that a red cell recovers the master sequence as it wanders aimlessly through sequence space is vanishingly small.  We can then treat the error rate $\mu$  as an irreversible mutation rate $ \mu = \mu_f$ which mutates green cells to red cells.  The red cells also acquire errors with rate $\mu$, but these errors will just move them around the flat fitness landscape.

The error threshold transition in the quasi-species model for the simple fitness landscape in Fig.~\ref{fig:quasispecies}  is identical to the directed percolation transition discussed in this paper.    In population genetics, these transitions are typically studied in well mixed populations, which corresponds to a mean field solution in the directed percolation language.  The mean field solution for the density of green cells $f(t)$ (fraction of cells with a master sequence) can be found from    Eq.~(\ref{eq:LangevinVMfull}) by assuming a homogeneous (space independent) density  $f(\mathbf{x},t) \equiv f(t)$ and negligible noise.  For an all green initial population $( f(t=0)  = 1)$, the fraction of green cells (averaged over many experiments) is given by the solution of $d \langle f(t) \rangle /dt=(s-\mu_f)\langle f(t) \rangle-s \langle f(t) \rangle^2:$
\begin{equation}
\langle f(t)\rangle  = 
\begin{cases}
\dfrac{s-\mu_f}{s-\mu_fe^{-t (s-\mu_f)} }  &  s \neq \mu_f \\[12pt]
 \dfrac{1}{1+ \mu_f t} & s = \mu_f
\end{cases}. \label{eq:MFf}
\end{equation}
Thus, $\langle f(t) \rangle$ approaches $0$ or $1 - \mu_f/s$ exponentially with time for $\mu_f > s$ and $\mu_f < s$, respectively.  The error threshold is given by the critical selective advantage $s_c = \mu_f$, and we evidently have a phase transition controlled by mutation-selection balance.   For more details about the relationship between quasi-species and mutation-selection balance, see Ref.~\cite{wilke}.  Note that this mean field analysis assumes a very large population size with negligible genetic drift -- the sharp phase transition is destroyed by number fluctuations in the well mixed ``zero-dimensional'' case.

 Finite spatial diffusion and noise due to genetic drift drastically change the  position of the phase transition and the broad features of the dynamics of $f(\mathbf{x},t)$.  Upon defining the rates $\mu_f$ and $s$ in units of the inverse cell division time (equivalent to setting the generation time $\tau_g = 1$), we construct the phase diagram (Fig.~\ref{fig:DKPhase}) by studying the limiting fraction of green cells $\lim_{t \rightarrow \infty} \langle f(\mathbf{x},t) \rangle_{\mathbf{x}}$ with the ``all green'' initial condition $f(\mathbf{x},0)=1$, where $\langle \ldots \rangle_{\mathbf{x}}$ is both an ensemble average and an average over all cell positions in the growing population at some time $t$.  The Monte Carlo simulations used to calculate this quantity are described in Sec.~\ref{SSimulations}.  A much stronger critical selective advantage ($s_c \sim \sqrt{\mu_f}$) than that expected from mean field theory $(s_c = \mu_f)$  is required to maintain the green strain in the population. These and other corrections to the mean field behavior have been extensively studied with both simulations and field theoretic methods \cite{Hinrichsen, NEQPTBook}. 

\section{\label{ACoarseGrain} Coarse-Graining and Langevin Description}

The simulations in this paper focus on voter-type models, with one individual per lattice site.  However, a coarse-graining procedure leads to a description in terms of continuous variables, similar to that reviewed for stepping stone models in Ref.~\cite{KorolevRMP}.  We coarse-grain voter-type models by considering a set of $\Omega > 1$ cells, and rescaling our lattice constant so that these $\Omega$ cells are all contained within a spatial volume $a^d$.   Following, e.g., Vazquez and L\'opez \cite{vazquez}, a coarse-grained population density of green cells is given by 
\begin{equation}
f(\mathbf{x},t) \equiv \frac{1}{\Omega} \sum_{i \in \mathcal{B}_{\Omega}(\mathbf{x})}  \frac{1+\sigma_i(t)}{2},
\end{equation}
where the sum is over cells in a neighborhood $\mathcal{B}_{\Omega}(\mathbf{x})$ of the $\Omega$ cells centered on $\mathbf{x}$, and we use Ising spin variables $\sigma_i(t) = \pm 1$ to specify the occupancy of site $i$.  Individual cells in this neighborhood change states according to  the rates given by Eq.~(\ref{eq:merates}).    We now randomly choose one of the $\Omega$ spins to flip in each neighborhood.
This means that each spin flip at $\mathbf{x}'$ changes our  field $f(\mathbf{x},t)$ by a displacement field $\pm r(\mathbf{x},\mathbf{x}')$ given by
\begin{equation}
 \pm r(\mathbf{x},\mathbf{x}') \equiv\pm     \Omega^{-1} a^{d}\delta(\mathbf{x}- \mathbf{x}'),
\end{equation}  
where the sign depends on the state of the spin at $\mathbf{x}'$.  In addition, we now use the average value of the spins in neighboring cells to evaluate the summations over nearest neighbors in the flipping rates due to selection and genetic drift (see Eq.~(\ref{eq:mutfliprates}) and Eq.~(\ref{eq:selfliprates})).  The resulting dynamics resembles the stepping stone model of population genetics \cite{KorolevRMP, kimura}, with populations of size $\Omega$ on compact ``islands'' where mutation and selection act, with exchange of individuals between neighboring islands. 

To study the dynamics of the the coarse-grained population density $f(\mathbf{x},t)$, we introduce rates    $\omega(f(\mathbf{x}); \pm r(\mathbf{x},\mathbf{x}'))$ that describe a transition from a starting field configuration $f$ to a field  $f\pm r$.  The new coarse-grained transition rates (in units of $\tau_g^{-1}$) that define an underlying master equation for the field $f$ are
\begin{align}
& \omega[f;+r] \equiv \omega \left[f(\mathbf{x}) \rightarrow f(\mathbf{x})+ r(\mathbf{x},\mathbf{x}') \right]\nonumber  \\[6pt]
  & \quad = [1-f(\mathbf{x})] \left  [ \mu_b + \frac{ 1}{z }\sum_{i \, \,  }  f(\mathbf{x}+\bm{\delta}_i)\right], \label{eq:continuousVMrates}
\end{align}
and
\begin{align}
&  \omega[f;-r] =  f(\mathbf{x}) \left  [ \vphantom{\sum_i} \mu_f +(1-s) \right. \nonumber \\
& \qquad \qquad \qquad \qquad \qquad\left. + \frac{  ^{}s^{}-1   }{z}\sum_{i \, \,  }  f(\mathbf{x}+\bm{\delta}_i)\right], \label{eq:continuousVMrates2}
\end{align}
where we sum over $z$ neighboring unit cells at vector distances $ \bm{\delta}_i$ from the cell centered on $\mathbf{x}$. For a cubic lattice $\bm{\delta}_i = a \hat{\mathbf{x}}_i$, where $\hat{\mathbf{x}}_i$ are Cartesian unit vectors.  

 To construct the master equation for the probability distribution functional $P[f,t]$ of the coarse-grained field,  we expand in $1/\Omega$.  This Kramers-Moyal expansion \cite{gardiner, vankampen, risken} leads to a Fokker-Planck equation (or ``Kolmogorov forward equation'') for the voter model dynamics. The rates $\omega(f; \pm r)$ will vary rapidly with the displacement $r$.  However, if  we choose  a $\Omega$ large enough, we can expand   $\omega(f; \pm r)$ around $r=0$ in powers of $\Omega^{-1}$ \cite{vankampen}.   To second order in $\Omega^{-1}$, the master equation for $P \equiv P[f(\mathbf{x}),t]$ follows from
\begin{align}
& \partial_t P=\int \mathrm{d}^{\mathrm{d}} \mathbf{x}' \left\{ \vphantom{\frac{1}{2}} \omega \left[  f+ r (\mathbf{x}');-r (\mathbf{x}') \right] P \left[ f+ r (\mathbf{x}')  ,t \right] \right.  \nonumber \\
 & \qquad \qquad   \, +\omega \left[  f-r (\mathbf{x}');r (\mathbf{x}') \right] P \left[ f-r (\mathbf{x}')  ,t \right]  \nonumber \\
& \qquad  \qquad  \left. -  \left[\omega \left(f ;   r (\mathbf{x}')\right) + \omega \left(f;-r(\mathbf{x}') \right)\right] P[f,t] \vphantom{\frac{1}{2}}\right\}, \end{align}
which leads to
\begin{align}
&\partial_t P   \approx -\frac{1}{\Omega}  \frac{\delta}{\delta f} \left\{  \left[\vphantom{\frac{1}{2}}\mu_b (1-f) -\mu_f f  \right.  \right. \nonumber \\[6pt]
& \qquad \qquad \left. \left.  + sf \left( 1-f \right)+ \frac{(1-sf)a^2}{z  } \nabla^2 f\right] P \vphantom{\frac{1}{2}}\right\} \nonumber \\
& \qquad +\frac{1}{2\Omega^2} \frac{\delta^2}{\delta f^2} \left\{\left[   \vphantom{\frac{1}{2}}\mu_{f}f+ \left( 2-s\right)f(1-f)  \right.\right. \nonumber \\
& \qquad \qquad \quad \left. \left.  +\mu_{b}(1-f)+ \frac{(s-f)a^2}{2z}  \, \nabla^2 f \right]P \right\} . \label{eq:FPVMfull}
\end{align}

Eq.~(\ref{eq:FPVMfull}) is a Fokker-Planck equation for the coarse-grained dynamics of the voter model in $d$-dimensions.  For a number of applications, it is convenient to shift to an equivalent Langevin description \cite{KorolevRMP}.
On setting $\Omega=1$  (to make contact with the lattice model) and assuming $s,\mu_{f},\mu_b \ll 1$,  we use the \^Ito prescription \cite{gardiner} to move from Eq.~(\ref{eq:FPVMfull}) to the stochastic differential equation for $f(\mathbf{x},t)$ given by Eq.~(\ref{eq:LangevinVMfull}) and Eq.~(\ref{eq:LangevinCorrVMfull})  in the main text.   The Langevin equation, when interpreted by the rules of the \^Ito calculus \cite{gardiner}, belongs to the universality class of directed percolation when $\mu_b = 0$ \cite{Hinrichsen}, i.e. for ``one-way'' mutations to the less fit phase in the presence of number fluctuations. Within an $\epsilon$-expansion about $d = 4$ spatial dimensions, the terms we have neglected are irrelevant in the renormalization group sense \cite{Hinrichsen}. Note also the equivalence to a stochastic stepping stone model    with one individual per deme \cite{KorolevRMP}.

Finally, we consider the connection between our radial expansion models and experiments.  Both the diffusion coefficient $D_r=a_r^2/z_r \tau_{g}$ and the front velocity $v$ could vary for different microbial radial expansions \cite{KorolevBac}.  Although the Langevin description in Eq.~(\ref{eq:RadLangevin}) and Eq.~(\ref{eq:RadLangevinnoise}) can accommodate this variation, our Bennett lattice model is constrained to have $D_r/(va_r) \approx a_r/(z_r v \tau_g)  \lesssim 1$. A likely  system that satisfies the bound is a \textit{Pseudomonas aeruginosa} radial expansion on a Petri dish \cite{KorolevBac}.   However, the bound will not be satisfied by all microbial radial expansions.

\section{\label{App:SingleSectorInfl} Single Sector Dynamics}

This appendix supplements the discussion in Sec.~\ref{SSOthers} by further developing the theory of the dynamics of a single green cell sector evolving in an otherwise red cell population under the radial Domany-Kinzel dynamics (see Sec.~\ref{SSimulations}) of a range expansion with a uniform circular front.  We will ignore mutations $(\mu_f=\mu_b=0)$ and allow for the green cells to enjoy a selective advantage $s$.    As discussed in the main text, the Langevin equation (Eq.~(\ref{eq:LangevinVMfull})) for the coarse-grained cell density $f(\mathbf{x},t)$ is difficult to analyze for nonzero $s$.  Hence,    we move to a ``dual'' description  of a single sector and consider the dynamics of  the two  sector boundaries at angular positions $\phi_1(t)$ and $\phi_2(t)$ at time $t$.   Without loss of generality, let $\phi_1(t)>\phi_2(t)$ so that $\phi_1(t)-\phi_2(t)$ is the angular sector size.  

Suppose that the dynamics of each boundary is independent, and that during each generation we either gain or lose a single cell at each boundary due to local competition with adjacent red cells.  Since the cells will generally be staggered along the population frontier, there will be slight variation in the change in the sector size when the sector loses or gains a cell.  Hence, we define a parameter $\tilde{a}_r$ equal to the average change in sector size when we add a single cell at the sector boundary.  The parameter $\tilde{a}_r$ should be close to the effective lattice spacing $a_r$.    The master equation for the probability $P(\phi_1,t)$ of observing a boundary at $\phi_1$ at time $t$  reads
\begin{align}
\partial_t P(\phi_1,t) & = \frac{p_G}{\tau_g} P\left(\phi_1-\frac{\tilde{a}_r}{R(t)},t\right) \nonumber \\
& \qquad \qquad -\left(\frac{1-p_G}{\tau_g}\right)P \left(\phi_1+\frac{\tilde{a}_r}{R(t)} ,t\right), \label{eq:MEsector}
\end{align}
where $\tau_g = 1$ is the generation time, and we approximate $ [ P(\phi,t+\tau_g)-P(\phi,t)]/\tau_g$ by  $\partial_t P(\phi,t) $.   The probability of a green cell out-competing a red cell is given by $p_G$.

 Recall that our Bennett model lattice is constructed so that the circular population radius advances by approximately one cell diameter per generation.  Thus, the angular position of the sector boundary at the frontier is determined by the color of approximately one cell.  If a single green cell competes with  adjacent red cells, the Domany-Kinzel rules (Eq.~(\ref{eq:pG})) imply that for $s \ll 1$,
\begin{equation}
p_G \approx  \frac{1}{2}- \frac{\gamma s}{4}, \label{eq:pGappx}
\end{equation}
where $\gamma = 1$ for simple competition between a pair of green and red cells.  The disorder in the lattice, however, may change $\gamma$ slightly as there can be a variable number of total parents $n_T$ that determine $p_G$ (see discussion of Eq.~(\ref{eq:pG}) in the main text). The continuum limit of the master equation Eq.~(\ref{eq:MEsector}) yields the Fokker-Planck equation, namely
\begin{align}
\partial_t P(\phi_1,t)&  =-  \frac{\gamma s \tilde{a}_r }{2 \tau_gR(t)} \, \partial_{\phi_1} P(\phi_1,t) \nonumber \\
& \quad + \frac{\tilde{a}_r^2}{2 \tau_g[R(t)]^2} \, \partial^2_{\phi_1}P(\phi_1,t). \label{eq:phi1FPE}
\end{align} 
Note that the other boundary position $\phi_2(t)$ also obeys Eq.~(\ref{eq:phi1FPE}).  The Fokker-Planck equations can be converted to stochastic differential equations for $\phi_1(t)$ and $\phi_2(t)$ using standard techniques \cite{gardiner}.  Subtracting the  stochastic differential equation for $\phi_2(t)$  from the one for $\phi_1(t)$ yields
\begin{equation}
\frac{d \phi}{dt} = \frac{\gamma  \tilde{a}_r s}{ \tau_g R(t)} \, \phi(t)+\sqrt{\frac{ 2 \tilde{a}_r^2}{\tau_g[R(t)]^2} \, } \, \eta(t), \label{eq:secSDE}
\end{equation}
where $\phi(t) = \phi_1(t)-\phi_2(t)$ is the angular sector size and $\eta(t) $ is a Gaussian white noise (with $\langle \eta(t) \eta(t') \rangle = \delta(t-t')$).    Upon defining the dimensionless conformal time $\tau \equiv t_c(t)/t^* = vt/(R_0+vt)=vt/R(t) $, we find that the Fokker-Planck equation associated with Eq.~(\ref{eq:secSDE})  (after changing variables from $t$ to $\tau$) must be the one given by Eq.~(\ref{eq:ssFPE}) in the main text, with
\begin{equation}
w = \frac{\gamma  \tilde{a}_rs}{v \tau_g} \mbox{\quad and \quad} \Delta = \frac{\tilde{a}_r^2}{R_0 v \tau_g}. \label{eq:wDelta}
\end{equation}

This simple model is not quite faithful to the simulations on an amorphous Bennett model lattice because the disorder in the lattice will introduce variability in the constant $\tilde{a}_r$ and the exact number of cells at the boundary of the sector per generation.  Some of these effects are modelled by $\gamma$, but we expect a slight $s$ dependence in $\Delta$, as well.  Also, the cell density of the Bennett model lattice decreases a little with increasing population radius.  However, by fitting the parameters $\tilde{a}_r$ and $\gamma$ to simulation results, we partially compensate for these complications, as confirmed by checks of the analytic results against our computer simulations.

\begin{figure}
\includegraphics[height=2.3in]{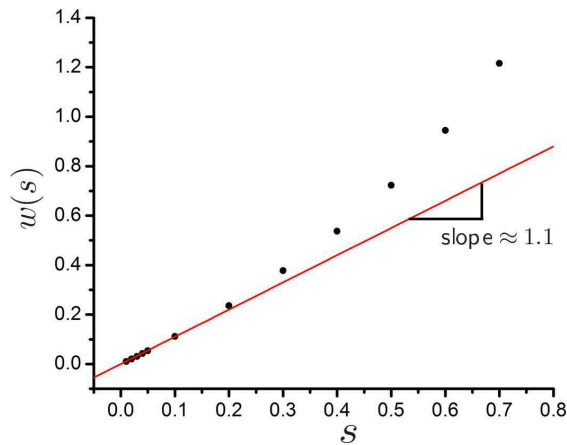}
\caption{\label{fig:gammaestimate}(Color online) The speed $w(s)$ of a green sector boundary at the frontier with various selection advantages $s$, invading an otherwise red population at the frontier.  The solid red line is the best fit of the speed to a linear function for $s \leq 0.1$ ($w(s) \approx 1.1 s$).  The initial angular sector size is $\phi_0 \approx \pi/2$, and the initial homeland radius $R_0 = 300$.  $w(s)$ is  calculated by averaging $(\langle \phi(t) \rangle - \phi_0)/\ln[R(t)/R_0]$ over approximately $10^3$ generations, where $\langle \phi(t) \rangle$ is the angular size of the green sector at time $t$, averaged over about $3 \times 10^4$  simulation runs. }
\end{figure}

When $s = 0$, Eq.~(\ref{eq:ssFPE}) in the main text reduces to a simple diffusion equation with diffusivity $\Delta=\tilde{a}_r^2/R_0 v \tau_g$, which can be solved exactly.  This solution is fitted to simulation results to find $\tilde{a}_r$.  We find $\tilde{a}_r \approx 0.91$ for $R_0 = 300$ (recall that $v = 1$ in the simulations).  Next, we pick a large initial sector angle $\phi_0 \approx \pi/2$ and measure the average angular size $\langle\phi(t) \rangle$    of the green sector as a function of time $t$.  Since the sector never goes extinct in this case, the dynamics of $\langle \phi(t) \rangle$ is controlled by the selection bias.  The green sector then forms, on average, the logarithmic spiral shape described by Eq.~(\ref{eq:determsector}) (see also Ref.~\cite{KorolevMueller}).  The shape of the spiral in Eq.~(\ref{eq:determsector}) implies that $w=(\langle \phi(t) \rangle - \phi_0)/ (\ln R(t) - \ln R_0)$ is a time independent constant.  The slope of the plot of $w$ versus $s$ yields an estimate for $\gamma  \tilde{a}_r/v$.    As expected from our simple model, we find in Fig.~\ref{fig:gammaestimate} a linear relationship between $w$ and $s$ for  $s \lesssim 1 $ that yields $\gamma \approx 1.2$.

\bibliographystyle{apsrev}
\bibliography{RadDKBib}

\begin{thebibliography}{50}
\expandafter\ifx\csname natexlab\endcsname\relax\def\natexlab#1{#1}\fi
\expandafter\ifx\csname bibnamefont\endcsname\relax
  \def\bibnamefont#1{#1}\fi
\expandafter\ifx\csname bibfnamefont\endcsname\relax
  \def\bibfnamefont#1{#1}\fi
\expandafter\ifx\csname citenamefont\endcsname\relax
  \def\citenamefont#1{#1}\fi
\expandafter\ifx\csname url\endcsname\relax
  \def\url#1{\texttt{#1}}\fi
\expandafter\ifx\csname urlprefix\endcsname\relax\def\urlprefix{URL }\fi
\providecommand{\bibinfo}[2]{#2}
\providecommand{\eprint}[2][]{\url{#2}}

\bibitem[{\citenamefont{Korolev et~al.}(2011)\citenamefont{Korolev, Xavier,
  Nelson, and Foster}}]{KorolevBac}
\bibinfo{author}{\bibfnamefont{K.~S.} \bibnamefont{Korolev}},
  \bibinfo{author}{\bibfnamefont{J.~B.} \bibnamefont{Xavier}},
  \bibinfo{author}{\bibfnamefont{D.~R.} \bibnamefont{Nelson}},
  \bibnamefont{and} \bibinfo{author}{\bibfnamefont{K.~R.}
  \bibnamefont{Foster}}, \bibinfo{journal}{The American Naturalist}
  \textbf{\bibinfo{volume}{178}}, \bibinfo{pages}{538} (\bibinfo{year}{2011}).

\bibitem[{\citenamefont{Merlo et~al.}(2006)\citenamefont{Merlo, Pepper, Reid,
  and Maley}}]{cancerEvo}
\bibinfo{author}{\bibfnamefont{L.~M.~F.} \bibnamefont{Merlo}},
  \bibinfo{author}{\bibfnamefont{J.~W.} \bibnamefont{Pepper}},
  \bibinfo{author}{\bibfnamefont{B.~J.} \bibnamefont{Reid}}, \bibnamefont{and}
  \bibinfo{author}{\bibfnamefont{C.~C.} \bibnamefont{Maley}},
  \bibinfo{journal}{Nat. Rev. Cancer} \textbf{\bibinfo{volume}{6}},
  \bibinfo{pages}{924} (\bibinfo{year}{2006}).

\bibitem[{\citenamefont{Cuesta et~al.}(2011)\citenamefont{Cuesta, Aguirre,
  Capit\'an, and Manrubia}}]{virus1}
\bibinfo{author}{\bibfnamefont{J.~A.} \bibnamefont{Cuesta}},
  \bibinfo{author}{\bibfnamefont{J.}~\bibnamefont{Aguirre}},
  \bibinfo{author}{\bibfnamefont{J.~A.} \bibnamefont{Capit\'an}},
  \bibnamefont{and} \bibinfo{author}{\bibfnamefont{S.~C.}
  \bibnamefont{Manrubia}}, \bibinfo{journal}{Phys. Rev. Lett.}
  \textbf{\bibinfo{volume}{106}}, \bibinfo{pages}{028104}
  (\bibinfo{year}{2011}).

\bibitem[{\citenamefont{Capit\'an et~al.}(2011)\citenamefont{Capit\'an, Cuesta,
  Manrubia, and Aguirre}}]{virus2}
\bibinfo{author}{\bibfnamefont{J.~A.} \bibnamefont{Capit\'an}},
  \bibinfo{author}{\bibfnamefont{J.~A.} \bibnamefont{Cuesta}},
  \bibinfo{author}{\bibfnamefont{S.~C.} \bibnamefont{Manrubia}},
  \bibnamefont{and} \bibinfo{author}{\bibfnamefont{J.}~\bibnamefont{Aguirre}},
  \bibinfo{journal}{PLoS ONE} \textbf{\bibinfo{volume}{6}},
  \bibinfo{pages}{e23358} (\bibinfo{year}{2011}).

\bibitem[{\citenamefont{Hewitt}(2000)}]{iceageEvo}
\bibinfo{author}{\bibfnamefont{G.}~\bibnamefont{Hewitt}},
  \bibinfo{journal}{Nature} \textbf{\bibinfo{volume}{405}},
  \bibinfo{pages}{907} (\bibinfo{year}{2000}).

\bibitem[{\citenamefont{Templeton}(2002)}]{humanmigration}
\bibinfo{author}{\bibfnamefont{A.}~\bibnamefont{Templeton}},
  \bibinfo{journal}{Nature} \textbf{\bibinfo{volume}{416}}, \bibinfo{pages}{45}
  (\bibinfo{year}{2002}).

\bibitem[{\citenamefont{Parmesan et~al.}(1999)\citenamefont{Parmesan, Ryrholm,
  Stefanescu, Hill, Thomas, Descimon, Huntley, Kaila, Kullberg, Tammaru
  et~al.}}]{climate1}
\bibinfo{author}{\bibfnamefont{C.}~\bibnamefont{Parmesan}},
  \bibinfo{author}{\bibfnamefont{N.}~\bibnamefont{Ryrholm}},
  \bibinfo{author}{\bibfnamefont{C.}~\bibnamefont{Stefanescu}},
  \bibinfo{author}{\bibfnamefont{J.~K.} \bibnamefont{Hill}},
  \bibinfo{author}{\bibfnamefont{C.~D.} \bibnamefont{Thomas}},
  \bibinfo{author}{\bibfnamefont{H.}~\bibnamefont{Descimon}},
  \bibinfo{author}{\bibfnamefont{B.}~\bibnamefont{Huntley}},
  \bibinfo{author}{\bibfnamefont{L.}~\bibnamefont{Kaila}},
  \bibinfo{author}{\bibfnamefont{J.}~\bibnamefont{Kullberg}},
  \bibinfo{author}{\bibfnamefont{T.}~\bibnamefont{Tammaru}},
  \bibnamefont{et~al.}, \bibinfo{journal}{Nature}
  \textbf{\bibinfo{volume}{399}}, \bibinfo{pages}{579} (\bibinfo{year}{1999}).

\bibitem[{\citenamefont{Hitch and Leberg}(2007)}]{climate2}
\bibinfo{author}{\bibfnamefont{A.~T.} \bibnamefont{Hitch}} \bibnamefont{and}
  \bibinfo{author}{\bibfnamefont{P.~L.} \bibnamefont{Leberg}},
  \bibinfo{journal}{Conservation Biology} \textbf{\bibinfo{volume}{21}},
  \bibinfo{pages}{534} (\bibinfo{year}{2007}).

\bibitem[{\citenamefont{Phillips et~al.}(2006)\citenamefont{Phillips, Brown,
  Webb, and Shine}}]{frogs}
\bibinfo{author}{\bibfnamefont{B.~J.} \bibnamefont{Phillips}},
  \bibinfo{author}{\bibfnamefont{G.~P.} \bibnamefont{Brown}},
  \bibinfo{author}{\bibfnamefont{J.~K.} \bibnamefont{Webb}}, \bibnamefont{and}
  \bibinfo{author}{\bibfnamefont{R.}~\bibnamefont{Shine}},
  \bibinfo{journal}{Nature} \textbf{\bibinfo{volume}{439}},
  \bibinfo{pages}{803} (\bibinfo{year}{2006}).

\bibitem[{\citenamefont{Gray et~al.}(2009)\citenamefont{Gray, Sappington,
  Miller, Moeser, and Bohn}}]{worms}
\bibinfo{author}{\bibfnamefont{M.~E.} \bibnamefont{Gray}},
  \bibinfo{author}{\bibfnamefont{T.~W.} \bibnamefont{Sappington}},
  \bibinfo{author}{\bibfnamefont{N.~J.} \bibnamefont{Miller}},
  \bibinfo{author}{\bibfnamefont{J.}~\bibnamefont{Moeser}}, \bibnamefont{and}
  \bibinfo{author}{\bibfnamefont{M.~O.} \bibnamefont{Bohn}},
  \bibinfo{journal}{Annu. Rev. Entomol.} \textbf{\bibinfo{volume}{54}},
  \bibinfo{pages}{303} (\bibinfo{year}{2009}).

\bibitem[{\citenamefont{Kimura and Weiss}(1964)}]{kimurapaper}
\bibinfo{author}{\bibfnamefont{M.}~\bibnamefont{Kimura}} \bibnamefont{and}
  \bibinfo{author}{\bibfnamefont{G.}~\bibnamefont{Weiss}},
  \bibinfo{journal}{Genetics} \textbf{\bibinfo{volume}{49}},
  \bibinfo{pages}{561} (\bibinfo{year}{1964}).

\bibitem[{\citenamefont{Korolev et~al.}(2010)\citenamefont{Korolev, Avlund,
  Hallatschek, and Nelson}}]{KorolevRMP}
\bibinfo{author}{\bibfnamefont{K.~S.} \bibnamefont{Korolev}},
  \bibinfo{author}{\bibfnamefont{M.}~\bibnamefont{Avlund}},
  \bibinfo{author}{\bibfnamefont{O.}~\bibnamefont{Hallatschek}},
  \bibnamefont{and} \bibinfo{author}{\bibfnamefont{D.~R.}
  \bibnamefont{Nelson}}, \bibinfo{journal}{Rev. Mod. Phys.}
  \textbf{\bibinfo{volume}{82}}, \bibinfo{pages}{1691} (\bibinfo{year}{2010}).

\bibitem[{\citenamefont{Moran}(1958)}]{Moran}
\bibinfo{author}{\bibfnamefont{P.~A.~P.} \bibnamefont{Moran}},
  \bibinfo{journal}{Math. Proc. of the Camb. Phil. Soc.}
  \textbf{\bibinfo{volume}{54}}, \bibinfo{pages}{60} (\bibinfo{year}{1958}).

\bibitem[{\citenamefont{Ewens}(2004)}]{Ewens}
\bibinfo{author}{\bibfnamefont{W.~J.} \bibnamefont{Ewens}},
  \emph{\bibinfo{title}{Mathematical Population Genetics}},
  vol.~\bibinfo{volume}{I} (\bibinfo{publisher}{Springer},
  \bibinfo{address}{New York}, \bibinfo{year}{2004}), \bibinfo{edition}{2nd}
  ed.

\bibitem[{\citenamefont{Murray}(2001)}]{murray}
\bibinfo{author}{\bibfnamefont{J.~D.} \bibnamefont{Murray}},
  \emph{\bibinfo{title}{Mathematical Biology}}, vol. \bibinfo{volume}{I - An
  Introduction} (\bibinfo{publisher}{Springer-Verlag},
  \bibinfo{address}{Berlin}, \bibinfo{year}{2001}).

\bibitem[{\citenamefont{Fisher}(1937)}]{fisher}
\bibinfo{author}{\bibfnamefont{R.~A.} \bibnamefont{Fisher}},
  \bibinfo{journal}{Annals of Eugenics} \textbf{\bibinfo{volume}{7}},
  \bibinfo{pages}{355} (\bibinfo{year}{1937}).

\bibitem[{\citenamefont{Korolev et~al.}(2012)\citenamefont{Korolev, M\"uller,
  Karohan, Murray, Hallatschek, and Nelson}}]{KorolevMueller}
\bibinfo{author}{\bibfnamefont{K.~S.} \bibnamefont{Korolev}},
  \bibinfo{author}{\bibfnamefont{M.~J.~I.} \bibnamefont{M\"uller}},
  \bibinfo{author}{\bibfnamefont{N.}~\bibnamefont{Karohan}},
  \bibinfo{author}{\bibfnamefont{A.~W.} \bibnamefont{Murray}},
  \bibinfo{author}{\bibfnamefont{O.}~\bibnamefont{Hallatschek}},
  \bibnamefont{and} \bibinfo{author}{\bibfnamefont{D.~R.}
  \bibnamefont{Nelson}}, \bibinfo{journal}{Physical Biology}
  \textbf{\bibinfo{volume}{9}}, \bibinfo{pages}{026008} (\bibinfo{year}{2012}).

\bibitem[{\citenamefont{Saito and Muller-Krumbhaar}(1995)}]{Saito}
\bibinfo{author}{\bibfnamefont{Y.}~\bibnamefont{Saito}} \bibnamefont{and}
  \bibinfo{author}{\bibfnamefont{H.}~\bibnamefont{Muller-Krumbhaar}},
  \bibinfo{journal}{Phys. Rev. Lett.} \textbf{\bibinfo{volume}{74}},
  \bibinfo{pages}{4325} (\bibinfo{year}{1995}).

\bibitem[{\citenamefont{Hallatschek
  et~al.}(2007{\natexlab{a}})\citenamefont{Hallatschek, Hersen, Ramanathan, and
  Nelson}}]{DRNPNAS}
\bibinfo{author}{\bibfnamefont{O.}~\bibnamefont{Hallatschek}},
  \bibinfo{author}{\bibfnamefont{P.}~\bibnamefont{Hersen}},
  \bibinfo{author}{\bibfnamefont{S.}~\bibnamefont{Ramanathan}},
  \bibnamefont{and} \bibinfo{author}{\bibfnamefont{D.~R.}
  \bibnamefont{Nelson}}, \bibinfo{journal}{Proc. Nat. Acad. Sci.}
  \textbf{\bibinfo{volume}{104}}, \bibinfo{pages}{19926}
  (\bibinfo{year}{2007}{\natexlab{a}}).

\bibitem[{\citenamefont{Hallatschek and Nelson}(2010)}]{nelsonhallatschek}
\bibinfo{author}{\bibfnamefont{O.}~\bibnamefont{Hallatschek}} \bibnamefont{and}
  \bibinfo{author}{\bibfnamefont{D.~R.} \bibnamefont{Nelson}},
  \bibinfo{journal}{Evolution} \textbf{\bibinfo{volume}{64}},
  \bibinfo{pages}{193} (\bibinfo{year}{2010}).

\bibitem[{\citenamefont{Kuhr et~al.}(2011)\citenamefont{Kuhr, Leisner, and
  Frey}}]{frey}
\bibinfo{author}{\bibfnamefont{J.-T.} \bibnamefont{Kuhr}},
  \bibinfo{author}{\bibfnamefont{M.}~\bibnamefont{Leisner}}, \bibnamefont{and}
  \bibinfo{author}{\bibfnamefont{E.}~\bibnamefont{Frey}}, \bibinfo{journal}{New
  J. Phys.} \textbf{\bibinfo{volume}{13}}, \bibinfo{pages}{113013}
  (\bibinfo{year}{2011}).

\bibitem[{\citenamefont{Hinrichsen}(2000)}]{Hinrichsen}
\bibinfo{author}{\bibfnamefont{H.}~\bibnamefont{Hinrichsen}},
  \bibinfo{journal}{Adv. in Phys.} \textbf{\bibinfo{volume}{49}},
  \bibinfo{pages}{815} (\bibinfo{year}{2000}).

\bibitem[{\citenamefont{Lynch et~al.}(1993)\citenamefont{Lynch, B\"urger,
  Butcher, and Gabriel}}]{mutmeltdown}
\bibinfo{author}{\bibfnamefont{M.}~\bibnamefont{Lynch}},
  \bibinfo{author}{\bibfnamefont{R.}~\bibnamefont{B\"urger}},
  \bibinfo{author}{\bibfnamefont{D.}~\bibnamefont{Butcher}}, \bibnamefont{and}
  \bibinfo{author}{\bibfnamefont{W.}~\bibnamefont{Gabriel}},
  \bibinfo{journal}{Journal of Heredity} \textbf{\bibinfo{volume}{84}},
  \bibinfo{pages}{339} (\bibinfo{year}{1993}).

\bibitem[{\citenamefont{Timofeeva and Coombes}(2004)}]{firedfire}
\bibinfo{author}{\bibfnamefont{Y.}~\bibnamefont{Timofeeva}} \bibnamefont{and}
  \bibinfo{author}{\bibfnamefont{S.}~\bibnamefont{Coombes}},
  \bibinfo{journal}{Phys. Rev. E} \textbf{\bibinfo{volume}{70}},
  \bibinfo{pages}{062901} (\bibinfo{year}{2004}).

\bibitem[{\citenamefont{Hallatschek
  et~al.}(2007{\natexlab{b}})\citenamefont{Hallatschek, Hersen, Ramanathan, and
  Nelson}}]{hallatschekPNAS}
\bibinfo{author}{\bibfnamefont{O.}~\bibnamefont{Hallatschek}},
  \bibinfo{author}{\bibfnamefont{P.}~\bibnamefont{Hersen}},
  \bibinfo{author}{\bibfnamefont{S.}~\bibnamefont{Ramanathan}},
  \bibnamefont{and} \bibinfo{author}{\bibfnamefont{D.~R.}
  \bibnamefont{Nelson}}, \bibinfo{journal}{PNAS}
  \textbf{\bibinfo{volume}{104}}, \bibinfo{pages}{19926}
  (\bibinfo{year}{2007}{\natexlab{b}}).

\bibitem[{\citenamefont{L\"ubeck}(2004)}]{finitesize}
\bibinfo{author}{\bibfnamefont{S.}~\bibnamefont{L\"ubeck}},
  \bibinfo{journal}{Int. J. Mod. Phys. B} \textbf{\bibinfo{volume}{18}},
  \bibinfo{pages}{3977} (\bibinfo{year}{2004}).

\bibitem[{\citenamefont{Domany and Kinzel}(1984)}]{domany}
\bibinfo{author}{\bibfnamefont{E.}~\bibnamefont{Domany}} \bibnamefont{and}
  \bibinfo{author}{\bibfnamefont{W.}~\bibnamefont{Kinzel}},
  \bibinfo{journal}{Phys. Rev. Lett.} \textbf{\bibinfo{volume}{53}},
  \bibinfo{pages}{311} (\bibinfo{year}{1984}).

\bibitem[{\citenamefont{Liggett}(1985)}]{voter1}
\bibinfo{author}{\bibfnamefont{T.~M.} \bibnamefont{Liggett}},
  \emph{\bibinfo{title}{Interacting Particle Systems}}
  (\bibinfo{publisher}{Springer-Verlag}, \bibinfo{address}{New York},
  \bibinfo{year}{1985}).

\bibitem[{\citenamefont{Redner}(2001)}]{redner}
\bibinfo{author}{\bibfnamefont{S.}~\bibnamefont{Redner}},
  \emph{\bibinfo{title}{A Guide to First-Passage Processes}}
  (\bibinfo{publisher}{Cambridge University Press},
  \bibinfo{address}{Cambridge}, \bibinfo{year}{2001}).

\bibitem[{\citenamefont{L\"ubeck}(2006)}]{lubeck}
\bibinfo{author}{\bibfnamefont{S.}~\bibnamefont{L\"ubeck}},
  \bibinfo{journal}{J. Stat. Mech.} p. \bibinfo{pages}{P09009}
  (\bibinfo{year}{2006}).

\bibitem[{\citenamefont{Henkel et~al.}(2008)\citenamefont{Henkel, Hinrichsen,
  and L\"ubeck}}]{NEQPTBook}
\bibinfo{author}{\bibfnamefont{M.}~\bibnamefont{Henkel}},
  \bibinfo{author}{\bibfnamefont{H.}~\bibnamefont{Hinrichsen}},
  \bibnamefont{and} \bibinfo{author}{\bibfnamefont{S.}~\bibnamefont{L\"ubeck}},
  \emph{\bibinfo{title}{Non-Equilibrium Phase Transitions}}, vol.
  \bibinfo{volume}{I - Absorbing Phase Transitions}
  (\bibinfo{publisher}{Springer Science}, \bibinfo{address}{The Netherlands},
  \bibinfo{year}{2008}).

\bibitem[{\citenamefont{Bennett}(1972)}]{Bennett}
\bibinfo{author}{\bibfnamefont{C.~H.} \bibnamefont{Bennett}},
  \bibinfo{journal}{J. App. Phys.} \textbf{\bibinfo{volume}{43}},
  \bibinfo{pages}{2727} (\bibinfo{year}{1972}).

\bibitem[{\citenamefont{Rubinstein and Nelson}(1982)}]{rubinstein}
\bibinfo{author}{\bibfnamefont{M.}~\bibnamefont{Rubinstein}} \bibnamefont{and}
  \bibinfo{author}{\bibfnamefont{D.~R.} \bibnamefont{Nelson}},
  \bibinfo{journal}{Phys. Rev. B} \textbf{\bibinfo{volume}{26}},
  \bibinfo{pages}{6254} (\bibinfo{year}{1982}).

\bibitem[{\citenamefont{Gardiner}(1985)}]{gardiner}
\bibinfo{author}{\bibfnamefont{C.~W.} \bibnamefont{Gardiner}},
  \emph{\bibinfo{title}{Handbook of Stochastic Methods}}
  (\bibinfo{publisher}{Springer-Verlag}, \bibinfo{address}{Berlin},
  \bibinfo{year}{1985}), \bibinfo{edition}{2nd} ed.

\bibitem[{\citenamefont{Glauber}(1963)}]{glauber}
\bibinfo{author}{\bibfnamefont{R.~J.} \bibnamefont{Glauber}},
  \bibinfo{journal}{J. Math. Phys.} \textbf{\bibinfo{volume}{4}}
  (\bibinfo{year}{1963}).

\bibitem[{\citenamefont{Palumbo et~al.}(1971)\citenamefont{Palumbo, Johnson,
  Rieck, and Witter}}]{palumbo}
\bibinfo{author}{\bibfnamefont{S.~A.} \bibnamefont{Palumbo}},
  \bibinfo{author}{\bibfnamefont{M.~G.} \bibnamefont{Johnson}},
  \bibinfo{author}{\bibfnamefont{V.~T.} \bibnamefont{Rieck}}, \bibnamefont{and}
  \bibinfo{author}{\bibfnamefont{L.~D.} \bibnamefont{Witter}},
  \bibinfo{journal}{J. Gen. Microbiol.} \textbf{\bibinfo{volume}{66}},
  \bibinfo{pages}{137} (\bibinfo{year}{1971}).

\bibitem[{\citenamefont{Gray and Kirwan}(1974)}]{gray}
\bibinfo{author}{\bibfnamefont{B.~F.} \bibnamefont{Gray}} \bibnamefont{and}
  \bibinfo{author}{\bibfnamefont{N.~A.} \bibnamefont{Kirwan}},
  \bibinfo{journal}{Biophys. Chem.} \textbf{\bibinfo{volume}{1}},
  \bibinfo{pages}{204} (\bibinfo{year}{1974}).

\bibitem[{\citenamefont{Neill}(1997)}]{diffgeom}
\bibinfo{author}{\bibfnamefont{B.~O.} \bibnamefont{Neill}},
  \emph{\bibinfo{title}{Elementary Differential Geometry}}
  (\bibinfo{publisher}{Academic Press}, \bibinfo{address}{San Diego},
  \bibinfo{year}{1997}), \bibinfo{edition}{2nd} ed.

\bibitem[{\citenamefont{Ali and Grosskinsky}(2010)}]{roughconformal}
\bibinfo{author}{\bibfnamefont{A.}~\bibnamefont{Ali}} \bibnamefont{and}
  \bibinfo{author}{\bibfnamefont{S.}~\bibnamefont{Grosskinsky}},
  \bibinfo{journal}{Adv. Complex Syst.} \textbf{\bibinfo{volume}{13}},
  \bibinfo{pages}{249} (\bibinfo{year}{2010}).

\bibitem[{\citenamefont{Gradshteyn and Ryzhik}(2007)}]{ryzhik}
\bibinfo{author}{\bibfnamefont{I.~S.} \bibnamefont{Gradshteyn}}
  \bibnamefont{and} \bibinfo{author}{\bibfnamefont{I.~M.}
  \bibnamefont{Ryzhik}}, \emph{\bibinfo{title}{Table of Integrals, Series, and
  Products}} (\bibinfo{publisher}{Academic Press}, \bibinfo{address}{Oxford},
  \bibinfo{year}{2007}), \bibinfo{edition}{7th} ed.

\bibitem[{\citenamefont{Saff and Snider}(2003)}]{complex}
\bibinfo{author}{\bibfnamefont{E.~B.} \bibnamefont{Saff}} \bibnamefont{and}
  \bibinfo{author}{\bibfnamefont{A.~D.} \bibnamefont{Snider}},
  \emph{\bibinfo{title}{Fundamentals of Complex Analysis with Applications to
  Engineering and Science}} (\bibinfo{publisher}{Pearson Education},
  \bibinfo{address}{New Jersey}, \bibinfo{year}{2003}), \bibinfo{edition}{3rd}
  ed.

\bibitem[{\citenamefont{Crow and Kimura}(1970)}]{kimura}
\bibinfo{author}{\bibfnamefont{J.~F.} \bibnamefont{Crow}} \bibnamefont{and}
  \bibinfo{author}{\bibfnamefont{M.}~\bibnamefont{Kimura}},
  \emph{\bibinfo{title}{An Introduction To Population Genetics Theory}}
  (\bibinfo{publisher}{Harper \& Row}, \bibinfo{address}{New York},
  \bibinfo{year}{1970}).

\bibitem[{\citenamefont{Eigen et~al.}(1988)\citenamefont{Eigen, McCaskill, and
  Schuster}}]{eigen}
\bibinfo{author}{\bibfnamefont{M.}~\bibnamefont{Eigen}},
  \bibinfo{author}{\bibfnamefont{J.}~\bibnamefont{McCaskill}},
  \bibnamefont{and} \bibinfo{author}{\bibfnamefont{P.}~\bibnamefont{Schuster}},
  \bibinfo{journal}{J. Phys. Chem.} \textbf{\bibinfo{volume}{92}},
  \bibinfo{pages}{6881} (\bibinfo{year}{1988}).

\bibitem[{\citenamefont{Jain and Krug}(2007)}]{quasispecies1}
\bibinfo{author}{\bibfnamefont{K.}~\bibnamefont{Jain}} \bibnamefont{and}
  \bibinfo{author}{\bibfnamefont{J.}~\bibnamefont{Krug}}, in
  \emph{\bibinfo{booktitle}{Structural Approaches to Sequence Evolution}},
  edited by \bibinfo{editor}{\bibfnamefont{U.}~\bibnamefont{Bastolla}},
  \bibinfo{editor}{\bibfnamefont{M.}~\bibnamefont{Porto}},
  \bibinfo{editor}{\bibfnamefont{H.~E.} \bibnamefont{Roman}}, \bibnamefont{and}
  \bibinfo{editor}{\bibfnamefont{M.}~\bibnamefont{Vendruscolo}}
  (\bibinfo{publisher}{Springer Berlin Heidelberg}, \bibinfo{year}{2007}),
  Biological and Medical Physics, Biomedical Engineering, pp.
  \bibinfo{pages}{299--339}.

\bibitem[{\citenamefont{Bull et~al.}(2005)\citenamefont{Bull, Meyers, and
  Lachmann}}]{quasispecies2}
\bibinfo{author}{\bibfnamefont{J.~J.} \bibnamefont{Bull}},
  \bibinfo{author}{\bibfnamefont{L.~A.} \bibnamefont{Meyers}},
  \bibnamefont{and} \bibinfo{author}{\bibfnamefont{M.}~\bibnamefont{Lachmann}},
  \bibinfo{journal}{PLoS Comput Biol} \textbf{\bibinfo{volume}{1}},
  \bibinfo{pages}{e61} (\bibinfo{year}{2005}).

\bibitem[{\citenamefont{Kearny and Majumdar}(2005)}]{kearney}
\bibinfo{author}{\bibfnamefont{M.~J.} \bibnamefont{Kearny}} \bibnamefont{and}
  \bibinfo{author}{\bibfnamefont{S.~N.} \bibnamefont{Majumdar}},
  \bibinfo{journal}{J. Phys. A: Math. Gen.} \textbf{\bibinfo{volume}{38}},
  \bibinfo{pages}{4097} (\bibinfo{year}{2005}).

\bibitem[{\citenamefont{Wilke}(2005)}]{wilke}
\bibinfo{author}{\bibfnamefont{C.~O.} \bibnamefont{Wilke}},
  \bibinfo{journal}{BMC Evol. Biol.} \textbf{\bibinfo{volume}{5}}
  (\bibinfo{year}{2005}).

\bibitem[{\citenamefont{Vazquez and L\'opez}(2008)}]{vazquez}
\bibinfo{author}{\bibfnamefont{F.}~\bibnamefont{Vazquez}} \bibnamefont{and}
  \bibinfo{author}{\bibfnamefont{C.}~\bibnamefont{L\'opez}},
  \bibinfo{journal}{Phys. Rev. E} \textbf{\bibinfo{volume}{78}},
  \bibinfo{pages}{061127} (\bibinfo{year}{2008}).

\bibitem[{\citenamefont{Kampen}(1992)}]{vankampen}
\bibinfo{author}{\bibfnamefont{N.~G.~V.} \bibnamefont{Kampen}},
  \emph{\bibinfo{title}{Stochastic Processes in Physics and Chemistry}}
  (\bibinfo{publisher}{Elsevier}, \bibinfo{address}{The Netherlands},
  \bibinfo{year}{1992}).

\bibitem[{\citenamefont{Risken}(1989)}]{risken}
\bibinfo{author}{\bibfnamefont{H.}~\bibnamefont{Risken}},
  \emph{\bibinfo{title}{The Fokker-Planck Equation: Methods of Solution and
  Applications}} (\bibinfo{publisher}{Springer-Verlag},
  \bibinfo{address}{Berlin}, \bibinfo{year}{1989}).

\end{thebibliography}

\end{document}